\documentclass[12pt]{article}
\usepackage{amsmath}
\usepackage{graphicx}
\usepackage{enumerate}
\usepackage{natbib}
\usepackage{url} %

\usepackage{xr}
\usepackage{amsthm,amssymb,amsfonts,bbm,framed}
\usepackage[group-separator={,}]{siunitx}
\usepackage[capitalise]{cleveref}
\usepackage[inline, shortlabels]{enumitem}
\setlist[enumerate]{
  itemjoin={{, }},
  itemjoin*={{, and }},
}
\usepackage{tikz}
\usetikzlibrary{bayesnet}
\usetikzlibrary{shapes,arrows,patterns,decorations.pathreplacing}
\newcommand{\V}[1]{\ensuremath{\boldsymbol{#1}}} %
\newcommand{\M}[1]{\ensuremath{#1}} %

\newcommand{\groupfigure}[1][]{%
  \begin{tikzpicture}

    \draw[]  (0, 0) rectangle  (3, 3);
    \draw[]  (0, 2) rectangle  (1, 3);
    \draw[]  (0, 1) rectangle  (1, 2);
    \draw[]  (0, 0) rectangle  (1, 1);
    \draw[]  (1, 1) rectangle  (2, 2);
    \draw[]  (1, 0) rectangle  (2, 1);
    \draw[]  (2, 0) rectangle  (3, 1);

    \draw[]  (4, 0) rectangle  (5, 1);
    \draw[]  (4, 1) rectangle  (5, 2);
    \draw[]  (4, 2) rectangle  (5, 3);

    \node[above, scale = 1.5] at (1.5,3) {$\M{\beta}_A$};
    \node[above, scale = 1.5] at (4.5,3) {$\M{\beta}_X$};

    \draw[thick, decorate, decoration = {brace}] (-0.3,0) -- (-0.3,3);
    \node[scale = 1.5] at (-0.7, 1.5)  {$n$};
    \draw[thick, decorate, decoration = {brace,mirror}] (0,-0.3) --  (3,-0.3);
    \node[scale = 1.5] at (1.5, -0.7)  {$n$};
    \draw[thick, decorate, decoration = {brace}]  (3.7,0) -- (3.7,3);
    \node[scale = 1.5] at (3.3, 1.5)  {$n$};
    \draw[thick, decorate, decoration = {brace,mirror}]  (4,-0.3) -- (5,-0.3);
    \node[scale = 1.5] at (4.5, -0.9)  {$d$};

    #1
  \end{tikzpicture}%
}

\begin{document}

  \title{Predicting Responses from Weighted Networks with Node Covariates in an Application to Neuroimaging}
  \author{Daniel Kessler\\
    Departments of Statistics and Psychiatry, University of Michigan\\
    and \\
    Keith Levin \\
    Department of Statistics, University of Wisconsin-Madison\\
    and \\
    Elizaveta Levina \\
    Department of Statistics, University of Michigan}
  \maketitle

\begin{abstract}
  We consider the setting where many networks are observed on a common node set,
  and each observation comprises edge weights of a network, covariates observed at each node, and an overall response.
  The goal is to use the edge weights and node covariates to predict the response while identifying an interpretable set of predictive features.
  Our motivating application is neuroimaging,
  where edge weights encode functional connectivity measured between brain regions,
  node covariates encode task activations at each brain region,
  and the response is disease status or score on a behavioral task.
  We propose an approach that constructs feature groups based on assumed community structure (naturally occurring in neuroimaging applications).
  We propose two feature grouping schemes that incorporate both edge weights and node covariates,
  and we derive algorithms for optimization using an overlapping group LASSO penalty.
  Empirical results on synthetic data show that our method, relative to competing approaches,
  has similar or improved prediction error along with superior support recovery,
  enabling a more interpretable and potentially more accurate understanding of the underlying process.
  We also apply the method to neuroimaging data from the Human Connectome Project.
  Our approach is widely applicable in neuroimaging where interpretability is highly desired.
\end{abstract}

\section{Introduction}
\label{sec:netcov-intro}

Predicting a response such as psychiatric disease status from the brain scan of an individual is an increasingly common task in human neuroimaging \citep{CALHOUN2017135,ARBABSHIRANI2017137,burgosMachineLearningClassification2020}.
For example, studies aim to predict individual phenotypes from ``resting state'' functional connectivity \citep{khoslaMachineLearningRestingstate2019}.
Many of these studies also acquire spatially-localized features (e.g., brain activation in response to a cognitive task) on the same participants.
The simultaneous use of multiple modalities obtained from brain imaging (e.g., both connectivity and activation during a task) offers an opportunity for better prediction and deeper understanding in how various characteristics of the brain affect the phenotype \citep{calhounMultimodalFusionBrain2016}.

This suggests a general statistical problem: modeling a response $y$ as a function of one or more network-valued predictors.  Networks contain information about connections (edges) between units of observation (nodes), and may also have additional node-level information available (node covariates).  In the neuroimaging application above, the nodes are locations in the brain, and edges correspond to connectivity between these locations.  Typical neuroimaging networks are undirected, and edge weights represent the strength of connectivity, though this varies with imaging modality.  The networks can be treated as having a common node set by mapping every participant's brain onto a common anatomical atlas \citep[e.g.,][]{powerFunctionalNetworkOrganization2011}, also known as a parcellation.  Node covariates may take the form of, e.g., activation in response to a cognitive task or gray matter volume, and the response $y$ is a participant-level variable such as a cognitive task score or disease status.

Parcellations of the brain provide not only a common atlas but also a partition of nodes into  ``brain systems,'' which we may think of as ``communities''. These may be based on domain knowledge or on prior application of community detection algorithms \citep{yeoOrganizationHumanCerebral2011,powerFunctionalNetworkOrganization2011}. Obtaining results at the level of brain systems rather than individual nodes aids interpretability and comparison across studies, and helps to balance power and spatial specificity \citep{nobleImprovingPowerFunctional2022}.
Popular methods, such as connectome predictive modeling  \citep{shenUsingConnectomebasedPredictive2017}  and brain basis set modeling \citep{sripadaBasicUnitsInterIndividual2019}, tend to vectorize edge weights to obtain a ``bag of features'' \citep{chungStatisticalConnectomics2021} for use in conventional supervised learning algorithms.  These approaches do not account for community structure, offer only ad hoc interpretation at that level of resolution and typically do not incorporate node covariates.
Domain-agnostic variable selection methods \citep[e.g., the LASSO][]{tibshiraniRegressionShrinkageSelection1996} similarly disregard network structure, but do serve as a predictive performance benchmark for the methods we aim to develop here.

The goal of this work is to predict a response from edge weights and node covariates while providing interpretability at the level of network communities.
We term this a ``network-aware'' approach, in contrast to the ``bag of features'' methods discussed above.
We call our method {\em NetCov}:~prediction from {\em net}works with node {\em cov}ariates.
NetCov imposes structured penalties that reflect communities, in contrast to methods that first aggregate connectivity among communities \citep[e.g.,][]{yuChildhoodTraumaHistory2019}, which may fail when the signal within a community is heterogeneous.
We propose two schemes depending on the mechanism believed to be involved, and derive an efficient algorithm based on overlapping group LASSO to obtain interpretable group-sparse solutions. 

Previous uses of group LASSO penalties in neuroimaging include \citet{shimizuProbabilisticDiagnosisUnderstanding2015a}, which used both group and sparse group LASSO (SGL), with voxels grouped based on brain region, in the classification of depression using task fMRI.
\citet{relion2019NetworkClassificationApplications} proposed a prediction framework using an SGL penalty where each node has all its incident edge weights grouped together and applied this to a Schizophrenia dataset.
\citet{richie-halfordMultidimensionalAnalysisDetection2021} also proposed an SGL-based method intended for use with diffusion-weighted magnetic resonance imaging where voxel-level features are grouped based on tissue tract.
To the best of our knowledge, NetCov is the first method to (a) construct groups that include both edge weights and node covariates, naturally spanning multiple imaging modalities, and (b) leverage community information in the construction of groups.

While our goal is to predict a response from an observed network, a related line of work approaches the converse problem: characterize networks from observed labels.
For example, \citet{tangSemiparametricTwoSampleHypothesis2017} proposed a method for testing the hypothesis that two networks are drawn from the same distribution and applied it to neural connectome graphs.
\citet{ginestetHypothesisTestingNetwork2017} developed Wald-like hypothesis tests for samples of networks analogous to classical one- and two-sample tests, and illustrated the method on functional connectivity data.
\citet{xiaMultiscaleNetworkRegression2020} introduced a ``multi-scale network regression'' model in which edge weights are predicted using phenotypes in a penalized model. 
In more recent work, \citet{kimGraphawareModelingBrain2021} modeled edge weights using a mixed-effects framework with a network-aware variance structure to model how networks change by condition.

The remainder of the paper is organized as follows.
\cref{sec:netcov-method} presents our model and the two feature grouping schemes.  Model fitting is presented in \cref{sec:netcov-procedure}.    
\cref{sec:netcov-numer-exper} presents numerical experiments assessing our approach and comparing it to other methods.
We then apply our approach to data from a large human neuroimaging study in \cref{sec:netcov-application}.
We close with a discussion of limitations and future work in \cref{sec:netcov-summary-discussion}.  

\section{The NetCov Model and Network-Aware Penalties}
\label{sec:netcov-method}
We start by fixing notation.
Let $N$ be the number of observations (e.g., participants, in the neuroimaging context), and let the data collected for each participant $i = 1, \dots, N$ be the triple $\left( \M{A}^{(i)}, \M{X}^{(i)}, y^{(i)} \right)$.
Here $\M{A}^{(i)}$ is the $n \times n$ signed and weighted adjacency matrix associated with the $i$-th observation, on a common set of nodes labeled $1, \dots, n$, with $\M{A}^{(i)}_{kl}$ representing the weight of the edge from node $k$ to node $l$ for participant $i$.
The matrix $\M{X}^{(i)} \in \mathbb{R}^{n \times d}$ contains node covariates for participant $i$, with the $k$-th row corresponding to the covariates of node $k$.
The response variable $y^{(i)}$ for participant $i$ may be real-valued or categorical.
Finally, let $c: \left\{ 1,\ldots, n \right\} \rightarrow \left\{1,\ldots, K\right\}$ be the map that assigns each node $k$, $k= 1,\ldots, n$ to one of $K$ possible communities.
We assume this map is known (or learned previously) and will use it to construct feature groups.
Importantly, the response $y^{(i)}$, edge weights $\M{A}^{(i)}$, and node covariates $\M{X}^{(i)}$ are all observed.
While there may be correlations among them, they are not inferred one from the other.
The class of matrices $\M{A}^{(i)}$ may be further restricted based on the application at hand. For instance, in all our examples, the networks are undirected, and so the $\M{A}^{(i)}$ are symmetric matrices.

\subsection{Prediction Model}
\label{sec:netcov-model}

Since we desire interpretable parameters, we use a standard generalized linear model  \citep{mccullaghGeneralizedLinearModels1998} to relate the response $y$ to the predictors $\left( \M{A}, \M{X} \right)$.
Conditional on $\left( \M{A}, \M{X} \right)$, $y$ follows a distribution satisfying
\begin{equation*}
  \mathbb{E} \left[ y \right] =
  g^{-1} \left\{
    \mu
    + \operatorname{Trace} \left( \M{\beta}_{\M{A}}^{\intercal} \M{A} \right)
    + \operatorname{Trace} \left( \M{\beta}_{\M{X}}^{\intercal} \M{X} \right)
  \right\},
\end{equation*}
where $g^{-1}$ is the inverse of the link function $g$,  and $\mu \in \mathbb{R}$, $\M{\beta}_{\M{A}} \in \mathbb{R}^{n \times n}$, and $\M{\beta}_{\M{X}} \in \mathbb{R}^{n \times d}$ are (unknown) coefficients.
The choice of $g$ will depend on the setting: for continuous $y$ we may use the identity link function, and for binary $y$ letting $g^{-1}(t) = \frac{1}{1 + \operatorname{exp}(-t)}$ yields a logistic regression model.
Note that if the networks have no self-loops or are undirected, many of the entries of $\M{A}$ are either redundant or of no interest, and we can remove these terms from the model by constraining the corresponding entries of $\beta_{\M{A}}$ to also be zero.

It will often be notationally convenient to use $\V{Z}^{(i)} \in \mathbb{R}^p$ to denote an appropriately vectorized version of $\left( \M{A^{(i)}}, \M{X^{(i)}} \right)$ and to let $\V{\beta}  \in \mathbb{R}^p$, be an analogously vectorized version of $\left( \M{\beta}_{\M{A}}, \M{\beta}_{\M{X}} \right)$.  The dimension $p$ depends on the number of non-redundant entries; for instance, if the networks are undirected with no self-loops, then  $p = n(n-1)/2+ d n$.   If we write $\M{Z} \in \mathbb{R}^{N \times p}$ for the matrix with  $\V{Z}^{(i)}$ as the $i$-th row, we can write
$\mathbb{E} \left[ \V{y} \right] = g^{-1}\left( \mu + \M{Z} \V{\beta} \right),$
which substantially simplifies subsequent derivations.

\subsection{Feature Groups}
\label{sec:netcov-feature-grouping}
\begin{figure}[ht]
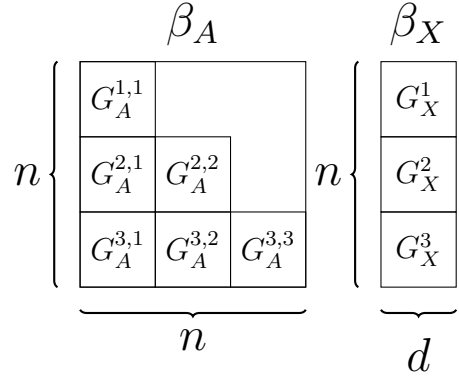

  \centering
  \groupfigure[{
    \node[] at (4.5,2.5) {$G_X^{1}$};
    \node[] at (4.5,1.5) {$G_X^{2}$};
    \node[] at (4.5,0.5) {$G_X^{3}$};
    \node[] at (0.5,2.5) {$G_A^{1, 1}$};
    \node[] at (0.5,1.5) {$G_A^{2, 1}$};
    \node[] at (0.5,0.5) {$G_A^{3, 1}$};
    \node[] at (1.5,1.5) {$G_A^{2, 2}$};
    \node[] at (1.5,0.5) {$G_A^{3, 2}$};
    \node[] at (2.5,0.5) {$G_A^{3, 3}$};
  }]
  \caption{Feature groups for undirected networks with $K=3$.  Network cells are on the left and node blocks are on the right.}
  \label{fig:grouping-primitives}
\end{figure}

Feature groups are at the core of our method and its goal to provide interpretable network-aware solutions. %
Suppose, without loss of generality, that the nodes are ordered such that community assignments are contiguous and non-decreasing, i.e., $i < j \implies c(i) \leq c(j)$.
Recall that coefficients corresponding to the node covariates are held in the matrix $\M{\beta}_{\M{X}} \in \mathbb{R}^{n \times d}$.
Since each row corresponds to coefficients associated with a distinct node, and each node is assigned to a community, we can partition the predictors in $\M{X}^{(i)}$ into $K$ ``blocks,'' where each block comprises the covariates associated with the nodes in a specific community.
Let $G_{\M{X}}^k = \left\{ i: c(i) = k \right\}$ denote a given block.  Recall that coefficients corresponding to edge weights are held in the matrix $\M{\beta}_{\M{A}} \in \mathbb{R}^{n \times n}$.
We can partition $\M{\beta}_{\M{A}}$ into ``cells,'' where each cell corresponds to coefficients associated with edges linking nodes belonging to a specific pair of communities.
If the network is directed, there will be $K^2$ cells, whereas if it is undirected there will be $K (K + 1)/2$ cells. %
Let $G_{\M{A}}^{k, k'} = \left\{ (i, j): c(i) = k \text{ and } c(j) = k' \right\}$ denote a given cell.
An example of this partitioning scheme is depicted in \cref{fig:grouping-primitives}.

\begin{figure}[ht]
  \centering
  \includegraphics[width=.3\textwidth]{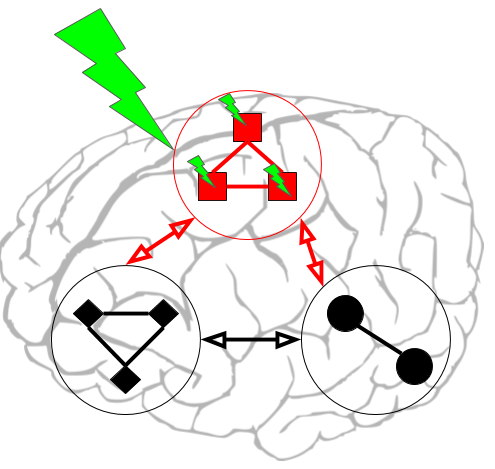} \qquad \vline \qquad
  \includegraphics[width=.3\textwidth]{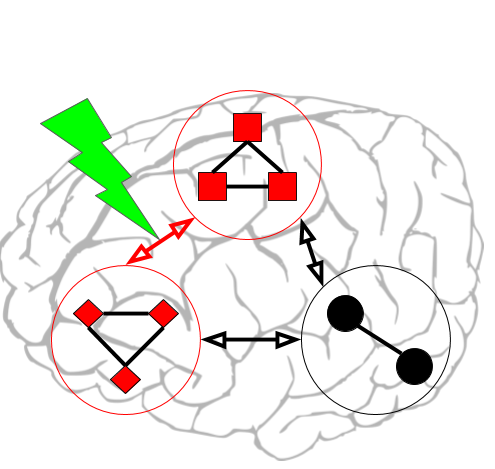}
  \caption{The neuroimaging motivation for grouping.  
    Circled groups of nodes represent brain systems, lines represent connectivity between nodes.
    Black is normal, red is abnormal, and a lightning bolt indicates a disease or injury.   Left (NBG):  a disease affects a system, and its connections to other systems become abnormal.   Right (EBG):  a disease affects connectivity between two systems, and the systems themselves become abnormal.}
  \label{fig:nbg-ebg-schematic}
\end{figure}

Blocks and cells form natural grouping units for nodal and edge covariates, respectively.
As such, we propose two feature grouping schemes: Node-Based Groups (NBG) and Edge-Based Groups (EBG).
These two schemes are motivated by the neuroscientific notion of ``lesion network mapping'' \citep{foxMappingSymptomsBrain2018}, and are schematically illustrated in \cref{fig:nbg-ebg-schematic}.
In the first scenario (NBG), an aberration (e.g., caused by disease, trauma, etc.) affects a brain system, corresponding to, say, community $k$.
This affects the block $G_{\M{X}}^{k}$ as well as its connectivity with other brain systems.
Thus, $G_{\M{A}}^{1, k}, G_{\M{A}}^{2, k}, \ldots, G_{\M{A}}^{K, k}$ are affected, too.
Formally, the $K$ feature groups ${\mathcal{G} = \left\{ G^{1}, G^{2},  \ldots, G^{K} \right\}}$ under NBG are given by 
\begin{equation*}
  G^{k} = G_{\M{X}}^k \cup \bigcup_{j = 1}^{K} G_{\M{A}}^{j, k}.
\end{equation*}

In the second scenario (EBG), an aberration affects edges instead of nodes, disrupting connectivity between two systems, say $k$ and $k'$, affecting $G_{\M{A}}^{k, k'}$.
This in turn affects covariates within both systems: $G_{\M{X}}^k$ and $G_{\M{X}}^{k'}$.
Formally, the $K(K+1)/2$ groups under EBG, ${\mathcal{G} = \left\{ G^{k, k'} : 1 \leq k \leq k' \leq K \right\}}$, are given by 
\begin{equation*}
  G^{k, k'} =   G_{\M{A}}^{k, k'} \cup G_{\M{X}}^k \cup G_{\M{X}}^{k'}.
\end{equation*}

\begin{figure}[ht]
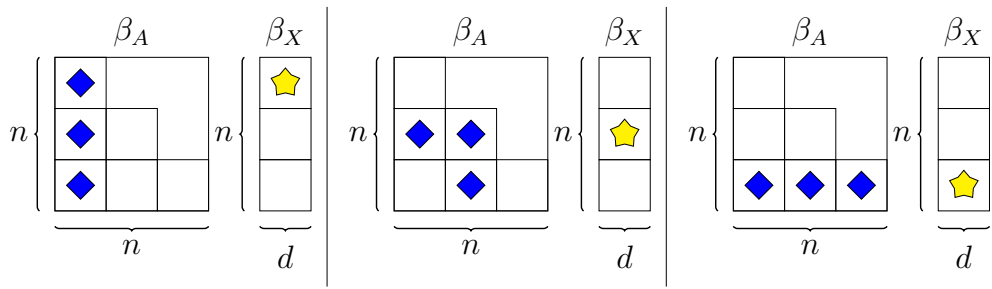

  \centering
  \resizebox{\textwidth}{!}{
    \begin{tabular}{c|c|c}
      \groupfigure[{
      \node[shape=star,draw,fill=yellow] at (4.5,2.5) {};
      \node[shape=diamond,draw,fill=blue] at (0.5,2.5) {};
      \node[shape=diamond,draw,fill=blue] at (0.5,1.5) {};
      \node[shape=diamond,draw,fill=blue] at (0.5,0.5) {};
      }]
      &
        \groupfigure[{
        \node[shape=star,draw,fill=yellow] at (4.5,1.5) {};
        \node[shape=diamond,draw,fill=blue] at (0.5,1.5) {};
        \node[shape=diamond,draw,fill=blue] at (1.5,1.5) {};
        \node[shape=diamond,draw,fill=blue] at (1.5,0.5) {};
        }]
      &
        \groupfigure[{
        \node[shape=star,draw,fill=yellow] at (4.5,0.5) {};
        \node[shape=diamond,draw,fill=blue] at (0.5,0.5) {};
        \node[shape=diamond,draw,fill=blue] at (1.5,0.5) {};
        \node[shape=diamond,draw,fill=blue] at (2.5,0.5) {};
        }]
    \end{tabular}
  }
  \caption{
    The NBG feature groups for $K=3$.  The panels, from left to right, show features associated with communities 1, 2, and 3.   Yellow stars correspond to node covariates, and blue diamonds to edge weights.
  }
  \label{fig:nbg}
\end{figure}

\begin{figure}[ht]
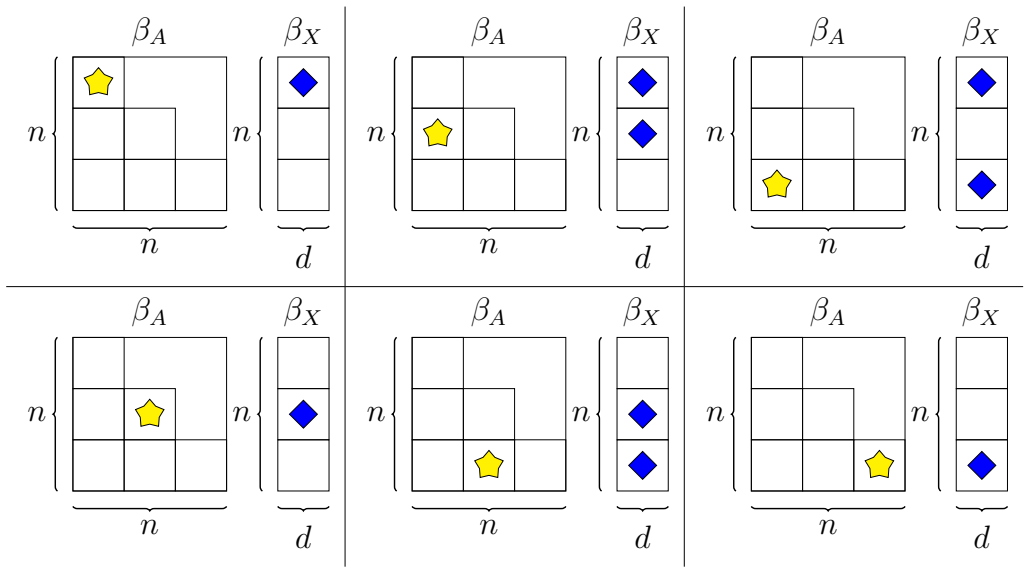

  \centering
  \resizebox{\textwidth}{!}{
    \begin{tabular}{c|c|c}
      \groupfigure[{
      \node[shape=star,draw,fill=yellow] at (0.5,2.5) {};
      \node[shape=diamond,draw,fill=blue] at (4.5,2.5) {};
      }]
      &
        \groupfigure[{
        \node[shape=star,draw,fill=yellow] at (0.5,1.5) {};
        \node[shape=diamond,draw,fill=blue] at (4.5,2.5) {};
        \node[shape=diamond,draw,fill=blue] at (4.5,1.5) {};
        }]
      &
        \groupfigure[{
        \node[shape=star,draw,fill=yellow] at (0.5,0.5) {};
        \node[shape=diamond,draw,fill=blue] at (4.5,2.5) {};
        \node[shape=diamond,draw,fill=blue] at (4.5,0.5) {};
        }]
      \\ \hline
      \groupfigure[{
      \node[shape=star,draw,fill=yellow] at (1.5,1.5) {};
      \node[shape=diamond,draw,fill=blue] at (4.5,1.5) {};
      }]
      &
        \groupfigure[{
        \node[shape=star,draw,fill=yellow] at (1.5,0.5) {};
        \node[shape=diamond,draw,fill=blue] at (4.5,1.5) {};
        \node[shape=diamond,draw,fill=blue] at (4.5,0.5) {};
        }]
      &
        \groupfigure[{
        \node[shape=star,draw,fill=yellow] at (2.5,0.5) {};
        \node[shape=diamond,draw,fill=blue] at (4.5,0.5) {};
        }]
    \end{tabular}
  }

  \caption{
    The EBG feature groups for $K=3$.  Each panel shows one group corresponding to a connection between communities $k_1$ and $k_2$, with $1 \leq k_1 \leq
    k_2 \le 3$.   Yellow stars correspond to node covariates, and blue diamonds to edge weights.
  }
  \label{fig:ebg}
\end{figure}

To summarize, an NBG group corresponds to all nodes in a community and all edges it is involved in, while an EBG group corresponds to a connection between two communities, including both communities' nodes and edges between them.
Under both NBG and EBG, each feature appears in at least one group, but the groups overlap.  
A given group comprises coefficients associated with both edge weights and node covariates.
The corresponding groupings of the coefficients of $\V{\beta}$ are illustrated, for $K = 3$,  in \cref{fig:nbg} for NBG and \cref{fig:ebg} for EBG.
Note that the stated definitions can be readily extended to directed settings.

\section{An Algorithm for Fitting NetCov}
\label{sec:netcov-procedure}

Recall that $\M{Z} \in \mathbb{R}^{N \times p}$ is a design matrix with row vectors corresponding to the  vectorization of $\left( \M{A}^{(i)}, \M{X}^{(i)} \right)$, where the dimension $p$ will vary depending on the number of nodes, node covariates, whether the network is directed, etc.
$\mu \in \mathbb{R}$ is an intercept and $\V{\beta} \in \mathbb{R}^p$ gives coefficients corresponding to the columns of $\M{Z}$.  We write $\V{\beta}_G$ to denote the subvector of $\V{\beta}$ whose coefficients correspond to features in group $G$.    Similarly, $\M{Z}_G$ denotes the submatrix formed by the columns of the design matrix $\M{Z}$ corresponding to group $G$.  

\subsection{The Objective Function}

We fit NetCov following a standard approach of minimizing a penalized loss function,
\begin{equation}
  \label{eq:objective}
  Q(\mu, \V{\beta} \mid \M{Z}, \V{y}) =
  \frac{1}{n}
  L(\mu, \V{\beta} \mid \M{Z}, \V{y})
  + \Omega(\V{\beta}),
\end{equation}
where $L$ is a loss function measuring the fit to training data and $\Omega$ is a regularization penalty, which we use to encourage group sparsity.  
We use deviance as the loss function, which depends on the assumed distribution of the response.
For example, for linear models (i.e., with the identity link function), we use the least squares loss function given by
\begin{equation*}
  \frac{1}{2}
  \sum_{i=1}^N \left(
    y^{(i)}
  - \mu
  - \V{\beta}^{\intercal} \V{Z}^{(i)}
  \right)^2.
\end{equation*}
For a binary response $y \in \left\{ 0, 1 \right\}$, we use the logistic regression loss %
\begin{equation*}
  -2 \sum_{i=1}^N \left[
  y^{(i)} \left( \mu + \V{\beta}^{\intercal} \V{Z}^{(i)}  \right)
  - \log \left\{
    1 + \exp \left( \mu + \V{\beta}^{\intercal} \V{Z}^{(i)} \right)
  \right\}
  \right].
\end{equation*}

We assume that each predictor has been standardized to have mean 0 and variance 1, and a continuous response $y$ is also standardized to have mean \num{0} and variance \num{1}.     The means and variances are learned from the training data only and are then used to normalize the test data, which, as a result, may not have exactly mean 0 and variance 1.   There is another important and less trivial standardization step, discussed further in \cref{sec:netcov-stand-group-lasso}.

Recall that $\mathcal{G}$ denotes the collection of groups, where $G \in \mathcal{G}$ is a set of indices for a given group of variables. The structure of $G$ is determined by the mechanism we are looking to find.        We assume that the coefficients $\V{\beta}$ are group-sparse: only a small subset of feature groups are ``active'' (i.e., contain non-zero coefficients).
This assumption is both pragmatic and principled: the number of edge features is $O(n^2)$ which for all realistic networks puts us in a high-dimensional regime where regularization is necessary.   At the same time, selecting features in groups yields interpretability at the level of communities (brain systems, in our application), rather than at the level of individual edges and/or nodes.   This is a desirable property for many applications, and especially so for neuroimaging, where domain knowledge generally exists at a resolution more compatible with large communities than specific brain coordinates, and where individual measurements are noisy and unreliable.

To select features in groups, we will employ the group LASSO penalty, which under appropriate conditions enjoys similar or superior theoretical properties relative to the standard LASSO \citep{nardiAsymptoticPropertiesGroup2008,huangBenefitGroupSparsity2010}.
The group penalty we use is of the form introduced by \citet{yuan2006ModelSelectionEstimation}, 
\begin{equation}
  \label{eq:g-lasso}
  \Omega(\V{\beta}) = \lambda \sum_{G \in \mathcal{G}} \sqrt{\lvert G \rvert} \lVert \V{\beta}_G \rVert_2,
\end{equation}
where $\lambda$ is a tuning parameter and $\lvert G \rvert$ is the cardinality of group $G$. %
This yields a geometry analogous to the LASSO \citep{tibshiraniRegressionShrinkageSelection1996}, with critical points at group-sparse solutions.

The penalty in \cref{eq:g-lasso} has a crucial property: the critical points correspond to solutions where ``inactive'' groups have all coefficients set to $0$.
This produces a set of groups $\mathcal{G}_I \subseteq \mathcal{G}$ such that $\forall G \in \mathcal{G}_I, \V{\beta}_G = \V{0}$, and a given feature can only be active if \emph{all} of the groups in which it participates are active.
This may be desirable in some settings \citep[see, e.g.,][]{relion2019NetworkClassificationApplications}, but our model calls for exactly the opposite: related nodes and edges are affected jointly, as our feature groups overlap.
This problem is discussed in \citet{jacobGroupLassoOverlap2009,obozinskiGroupLassoOverlaps2011}, and we follow their approach of duplicating variables to render the  groups non-overlapping, then using the standard group LASSO formulation on the expanded variable set.
That is, we construct, via concatenation, $\M{Z}^{\star} = \begin{bmatrix} \M{Z}_G : G \in \mathcal{G} \end{bmatrix}$ and $\V{\beta}^{\star} = \begin{bmatrix} \V{\beta}_G^{\intercal} : G \in \mathcal{G} \end{bmatrix}^{\intercal}$, and then minimize $Q(\mu, \V{\beta}^{\star} \mid \M{Z}^{\star}, \V{y})$.
For visualization and interpretation purposes, we can map an estimate of $\V{\beta}^{\star}$ back to the dimension of $\V{\beta}$ through summation: suppose that the $i$-th coordinate of $\V{\beta}$ appears in two groups and corresponds to coordinates $j$ and $j'$ of $\V{\beta}^{\star}$. Then we can find
$\left( \V{\beta} \right)_i = \left( \V{\beta}^{\star} \right)_j + \left( \V{\beta}^{\star} \right)_{j'}$.

\subsection{Standardizing within Groups}
\label{sec:netcov-stand-group-lasso}
It is common when fitting the LASSO to standardize the columns of the design matrix $\M{Z}$ to have mean 0 and variance 1. For the group LASSO, \citet{simonStandardizationGroupLasso2012} argue that the appropriate normalization involves not only centering and rescaling columns, but orthonormalizing the columns corresponding to each group.
As discussed in \citet{brehenyGroupDescentAlgorithms2015}, this orthonormalization yields a more straightforward and efficient algorithm, and is tantamount to penalizing the contribution of each group to the linear predictor.
This approach, called the ``groupwise prediction penalty'' in \citet{buhlmannStatisticsHighDimensionalData2011}, can be accomplished in practice by computing the SVD of each $\M{Z}^{\star}_G = U_{G} \Sigma_{G} V_G^T$, where we limit the decomposition to singular vectors corresponding to nonzero singular values.
We then construct a new design matrix comprising orthonormalized groups as $\tilde{\M{Z}}^{\star} = \begin{bmatrix} U_G : G \in \mathcal{G}
\end{bmatrix}$.
We use this quantity when minimizing \cref{eq:objective} in $\tilde{\V{\beta}}^{\star}$.
Note that groups of less than full column rank will have their penalty in \cref{eq:g-lasso} scaled based on the number of columns of their corresponding $U_G$, i.e., the rank of $Z_G$.
After obtaining an optimal $\tilde{\V{\beta}}^{\star}$, it is possible to invert both the orthonormalization and variable duplication to arrive at a solution that is parameterized by $\V{\beta}$, which we use for both prediction and interpretation. See \citet{brehenyGroupDescentAlgorithms2015,zengOverlappingGroupLogistic2016} for more details.

\subsection{Implementation and Parameter Tuning}
\label{sec:netcov-parameter-tuning}

Efficient algorithms for solving the non-overlapping standardized group LASSO penalty in the context of both linear and logistic regression are presented in \citet{brehenyGroupDescentAlgorithms2015} and available in the \texttt{grpreg} package in \texttt{R} \citep{rcoreteamLanguageEnvironmentStatistical}.
To use this approach in the overlapping case, we use the \texttt{grpregOverlap} \citep{zengOverlappingGroupLogistic2016} package: it manages variable duplication and depends heavily on \texttt{grpreg}.
Unfortunately, as of this writing, this package is no longer available from CRAN, but it is available from Github at \url{https://github.com/YaohuiZeng/grpregOverlap};
this version incorporates a number of improvements and fixes that we contributed in the course of our present work.

In practice, the tuning parameter $\lambda$ must be learned from data.
Following standard practice, we use cross-validation to choose $\lambda$ for both NetCov and the regular LASSO, which we use as a baseline comparison.  
We broadly follow the default settings for \texttt{glmnet} \citep{friedmanRegularizationPathsGeneralized2010}.
First, we identify the data-driven quantity $\lambda_{\text{max}}$, the smallest value of $\lambda$ for which the selected model is fully sparse.
Then, we set $\lambda_{\text{min}} = 0.05 \lambda_{\text{max}}$ and create a logarithmically-spaced grid of candidate values for $\lambda$ between $\lambda_{\text{min}}$ and $\lambda_{\text{max}}$.
We then conduct ten-fold cross-validation at each of the grid values, and compute average out-of-sample deviances.  
Letting $\lambda^{\star}$ be the grid value that minimizes the mean deviance, we set $\hat{\lambda}$ to the largest grid value whose out-of-sample deviance is within one standard error of that corresponding to $\lambda^{\star}$.    %
Finally, we refit the model to the full training set with $\lambda = \hat{\lambda}$. 

\section{Numerical Experiments}
\label{sec:netcov-numer-exper}

We conduct numerical experiments in a variety of settings to assess the performance of our procedure and to compare with competing strategies.  Generating simulated data involves generating or specifying the covariates $\M{A}^{(i)}$, $\M{X}^{(i)}$, specifying the coefficients $\V{\beta}$, and drawing the response $\V{y}$ from an appropriate model.  
In our first set of simulations in \cref{sec:netcov-experiment-i} the design matrix is synthetic, which allows us to vary more parameters and explore their influence on performance.
In \cref{sec:netcov-experiment-iii}, we fix the design to correspond to covariates from a human neuroimaging study.

To set coefficients $\V{\beta}$,  we vary
\begin{enumerate*}[(i), after={.}]
\item NBG vs EBG group structure
\item number of active feature groups (either \num{1} or \num{5})
\item the magnitude of active features (i.e., controlling the signal-to-noise ratio [SNR])
\end{enumerate*}
We simulate all these scenarios for both continuous and binary responses.
For simplicity, we only consider undirected networks with no self-loops and keep the number of nodal covariates $d=1$.
However, all results can be readily extended beyond these settings.

All simulations include both a training set used for both parameter tuning and model fitting and a test set used to assess out-of-sample performance.
While there are \num{8} unique combinations of group structure, number of active groups, and response (continuous versus binary),
we smoothly vary the SNR across \num{20} levels;
this leads to 160 unique settings for each experiment.
We repeat each experiment 10 times for each setting and average the results.
Because we conduct a total of nearly \num{10000} experiments;
we conduct our simulations on a high performance computing system %
with the extremely useful \texttt{R} package \texttt{batchtools} \citep{langBatchtoolsToolsWork2017}.

As a baseline comparison to our method, we include regular LASSO
as implemented in the \texttt{glmnet} \citep{friedmanRegularizationPathsGeneralized2010} package.
For both NetCov and LASSO, the tuning parameter is chosen by cross-validation as described in \cref{sec:netcov-parameter-tuning}.
For the simulations, we do not include other potential competing methods developed specifically for neuroimaging, such as Brain Basis Sets \citep{sripadaBasicUnitsInterIndividual2019} because they do not perform feature selection.

Since our primary goal is interpretation obtained from variable selection, we look at  support recovery as a measure of performance, computing both recall and precision for $\V{\beta}$,
where recall is defined as
\begin{equation*}
  \frac{\operatorname{TP}}{\operatorname{TP} + \operatorname{FN}},
\end{equation*}
and precision is defined as
\begin{equation*}
  \frac{\operatorname{TP}}{\operatorname{TP} + \operatorname{FP}},
\end{equation*}
where $\operatorname{TP}$ denotes the number of true nonzero coefficients that are in the estimated support, $\operatorname{FP}$ denotes the number of true zero coefficients that are in the estimated support, and $\operatorname{FN}$ denotes the number of true nonzero coefficients that are not in the estimated support.

We also assess out-of-sample prediction performance using out-of-sample classification accuracy for binary responses and using the correlation between our predictions $\hat{\V{y}}$ and the observed values in the test set, for the ease of comparison with the neuroimaging literature, which uses this measure (e.g., \citet{sripadaTreadmillTestCognition2020,hsuRestingstateFunctionalConnectivity2018}).   We plot these metrics against SNR for linear models and against Bayes Error (BE) for logistic models, given by
\begin{equation}
  \label{eq:snr}
  \operatorname{SNR} = \sigma^{-2} \operatorname{Var}_{\V{Z}} \left( \V{Z} \V{\beta} \right)
\end{equation}
and
\begin{equation}
  \label{eq:bayes-error}
  \operatorname{BE} = E_{\V{Z}} \left[  \operatorname{min} \left(\operatorname{logit}^{-1} \left( \V{Z} \V{\beta} \right),  1 - \operatorname{logit}^{-1} \left( \V{Z} \V{\beta} \right)  \right)  \right].
\end{equation}
In all of our simulations with continuous responses,
we set the error variance $\sigma^2 = 1$.
When working with semi-synthetic data, where the design is based on real data,  as discussed in \cref{sec:netcov-experiment-iii}, we compute the expectation and variance with respect to the empirical distribution of the training data.

\subsection{Experiment I: Fully Synthetic Data}
\label{sec:netcov-experiment-i}
In this experiment, we generate fully synthetic designs $\left( \M{A}^{(i)}, \M{X}^{(i)} \right)_{i=1}^N$.
We set the number of observations $N =$ \num{1000} and the number of communities $K =$ \num{10}, with 5 nodes each for a total of  $n=50$ nodes.   These are on the order of what is expected in neuroimaging settings, albeit on the low end, to accommodate running a large number of simulations.   
All unique entries of $\M{A}^{(i)}$ and $\M{X}^{(i)}$ are drawn independently from a standard normal distribution.   %

We specify $\V{\beta}$ by first fixing its support and then setting all nonzero entries to the same constant $\alpha$, which will be varied to control SNR.  
We form feature groups according to either the NBG or EBG scheme as described in \cref{sec:netcov-feature-grouping}.  
We then select either one or five groups to be active and take their union to be the active feature set.
For NBG, the selected groups are $\left\{ G^{(1)} \right\}$ or $\left\{ G^{(1)}, G^{(2)}, G^{(3)}, G^{(4)}, G^{(5)}  \right\}$.
For EBG, the selected groups are $\left\{ G^{(1, 1)} \right\}$ or $\left\{ G^{(1, 1)}, G^{(3, 1)}, G^{(3, 2)}, G^{(4, 4)}, G^{(6, 5)} \right\}$.  For EBG, we chose these five to include both diagonal and off-diagonal cells and to vary the amount of overlap.
We always set the intercept $\mu = 0$.
Finally, we draw each $y^{(i)}$ according to either a linear model $y^{(i)} = \V{Z}^{(i)} \V{\beta} + \epsilon^{(i)}$, where each $\epsilon^{(i)}$ is  an independent standard normal, or a logistic model, where each $y^{(i)}$ is an independent Bernoulli random variable with success probability $\operatorname{logit}^{-1} \left( \V{Z}^{(i)} \V{\beta} \right)$.
For each realization, the training and the test set have the same design matrix $Z$ and $\V{\beta}$, and differ only in their responses $\V{y}$, which are drawn independently.  NetCov is fit with the true group structure, since we treat it as known.

\begin{figure}[ht]
  \centering
  \includegraphics[width=\textwidth]{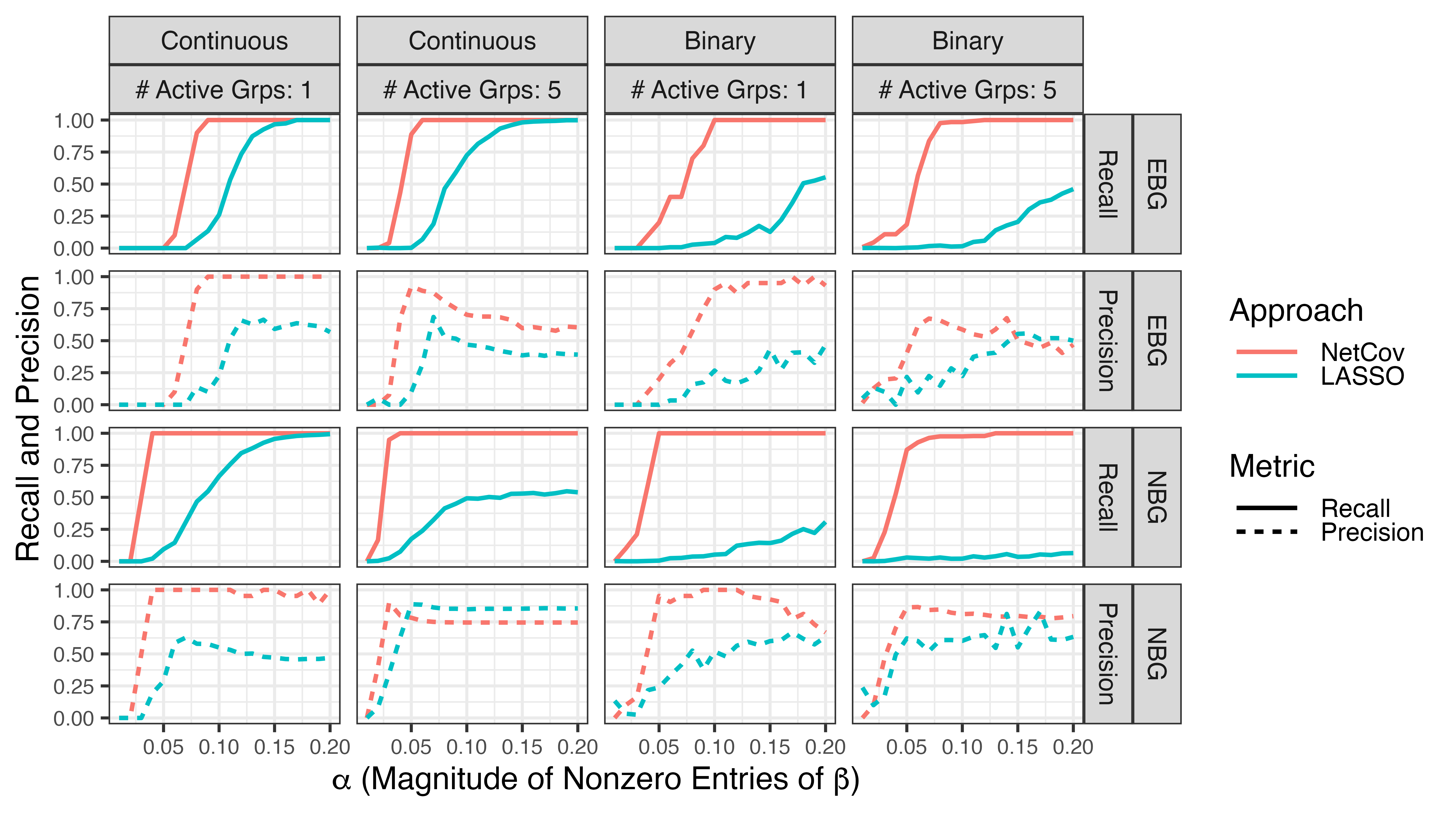}
  \caption{Support recovery in Experiment I: recall and precision as a function of nonzero coefficient magnitude $\alpha$ for NetCov (red) and LASSO (blue).  Each of the four columns corresponds to either continuous or binary response and either 1 or 5 active groups.  Each of the four rows corresponds to either EBG or NBG and either support recovery or precision for $\V{\beta}$.
}
  \label{fig:exp-I-support}
\end{figure}

Results in \cref{fig:exp-I-support} show that, 
as expected, support recovery generally improves with increasing SNR, and NetCov yields superior support recovery, especially on recall in low and intermediate SNR regimes.
NetCov is also generally superior to the LASSO on precision, although in high SNR settings the pattern is sometimes decreasing.
While this at first may seem surprising, this is a result of the parameter tuning process.
Both NetCov and LASSO incur bias due to the use of penalization.
There is a tendency for cross-validation to choose a small value of $\lambda$ in order to reduce this bias, but this comes at the cost of selecting inactive features which harms precision (but not recall).
In \cref{sec:netcov-roc-curves}, the receiver operating characteristic curves that characterize behavior along the entire $\lambda$ path show that NetCov:EBG generally dominates the LASSO.  
See also \citet{wangPriceCompetitionEffect2020} for a discussion of circumstances in which growing signal strength does not yield improved support recovery, chiefly due to effect size heterogeneity.  

\begin{figure}[ht]
  \centering
  \includegraphics[width=\textwidth]{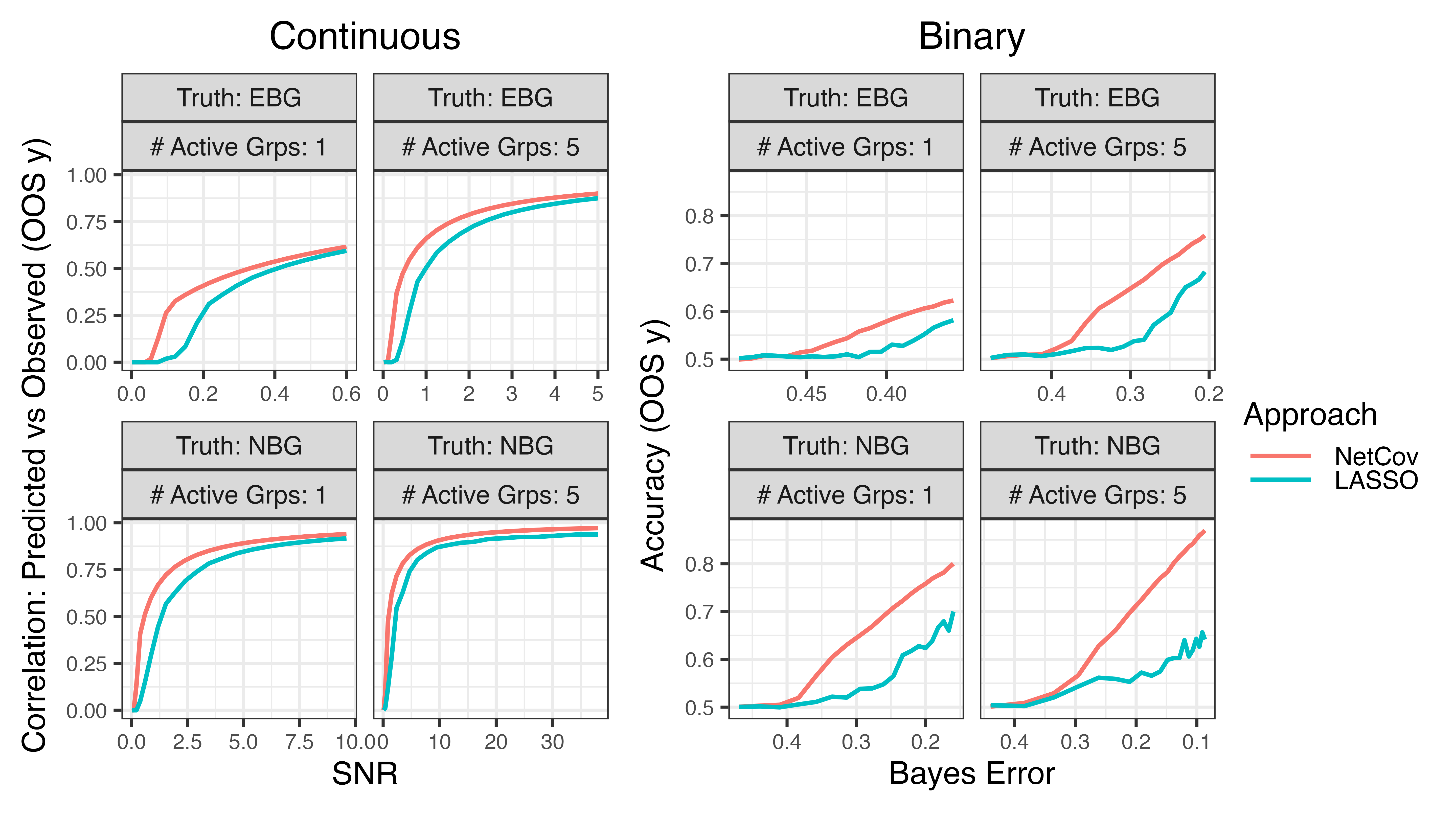}
  \caption{
    Out-of-sample prediction performance in Experiment I as a function of problem difficulty (SNR for continuous response and Bayes error for binary) for NetCov (red) and LASSO (blue).  Note the horizontal scale is different in every panel
    }
  \label{fig:exp-I-pred}
\end{figure}

Out-of-sample prediction  performance in the continuous and binary cases is shown  in \cref{fig:exp-I-pred}.  
Consistent with the improved support recovery, out-of-sample prediction for NetCov is generally superior to the LASSO.    This suggests that when the additional structure imposed by NetCov is present in the data, NetCov will yield not only superior support recovery and interpretability relative to the LASSO, but also improved out-of-sample prediction performance.     This improvement comes at a cost of trading off additional flexibility of the LASSO, which is an advantage in less structured models.  

\subsection{Experiment II: Semi-Synthetic Data}
\label{sec:netcov-experiment-iii}

In this experiment, we use data from the neuroimaging application as the design matrix; see \cref{sec:netcov-application} for more details regarding this dataset.   We have \num{785} observations in the training set and \num{96} observations in the test set.   
In addition to signed, weighted edges in the adjacency matrices, we also have a single continuous covariate associated with each node.  The edge weights represent functional connectivity in resting state fMRI and the covariate is measured during a working memory task-based fMRI session;  see \cref{sec:netcov-application} for details.   
Like in Experiment I, these networks are undirected without self-loops.  %

In order to keep the intercept at 0 as in Experiment I, we center the columns of the training design matrix and subtract these column means from the test design matrix.
In all other respects, this experiment is identical to Experiment I in \cref{sec:netcov-experiment-i}, where we specify the support of $\V{\beta}$ to involve either one or five groups under either the EBG of NBG schemes.   This semi-synthetic experiment involves real-world data with unknown dependence structure for $A$ and $X$, but we draw responses $\V{y}$ from our model with known $\V{\beta}$, which allows us to assess support recovery and other performance metrics.

Each network in the sample has \num{236} nodes, and each node can be assigned to one of \num{13} communities.
The names of these communities and the number of nodes in each are given in
Supplementary Table \ref{tab:power-communities}.
However, using this community assignment scheme results in many feature groups that have more predictors than we have observations,
and this is problematic for the group LASSO
(see Appendix \ref{sec:netcov-experiment-ii} for more discussion).
As a simple remedy, in this experiment we randomly break up the large communities into smaller pieces until we arrive at a parcellation that has 50 communities where \num{15} of the communities have \num{4} nodes, \num{34} communities have \num{5} nodes, and a single community has \num{6} nodes.
This modification does not change the overall covariance structure of $\M{Z}$, but it does change the intra-group covariance structure and also puts all groups on roughly equal footing in terms of size.
We present results when the original community assignments are used in Appendix \ref{sec:netcov-experiment-ii}.

\begin{figure}[ht]
  \centering
  \includegraphics[width=\textwidth]{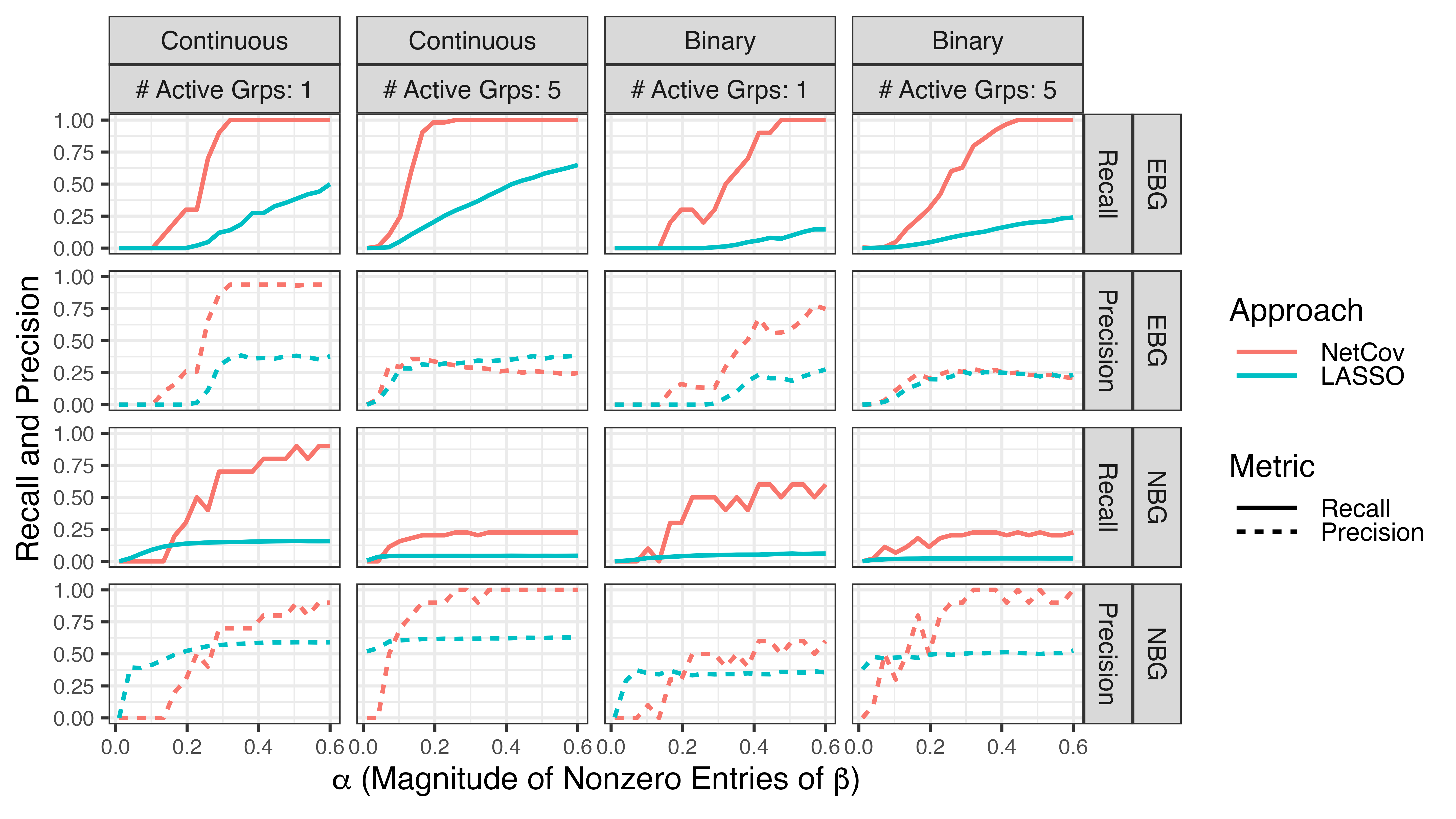}
  \caption{Support recovery (as measured by recall and precision) in Experiment II by NetCov (red) and LASSO (blue) as a function of nonzero coefficient magnitude $\alpha$. Each of the four columns corresponds to either continuous or binary response and either 1 or 5 active groups.  Each of the four rows corresponds to either EBG or NBG and either support recovery or precision for $\V{\beta}$.
  }
  \label{fig:exp-III-support}
\end{figure}

\begin{figure}[ht]
  \centering
  \includegraphics[width=\textwidth]{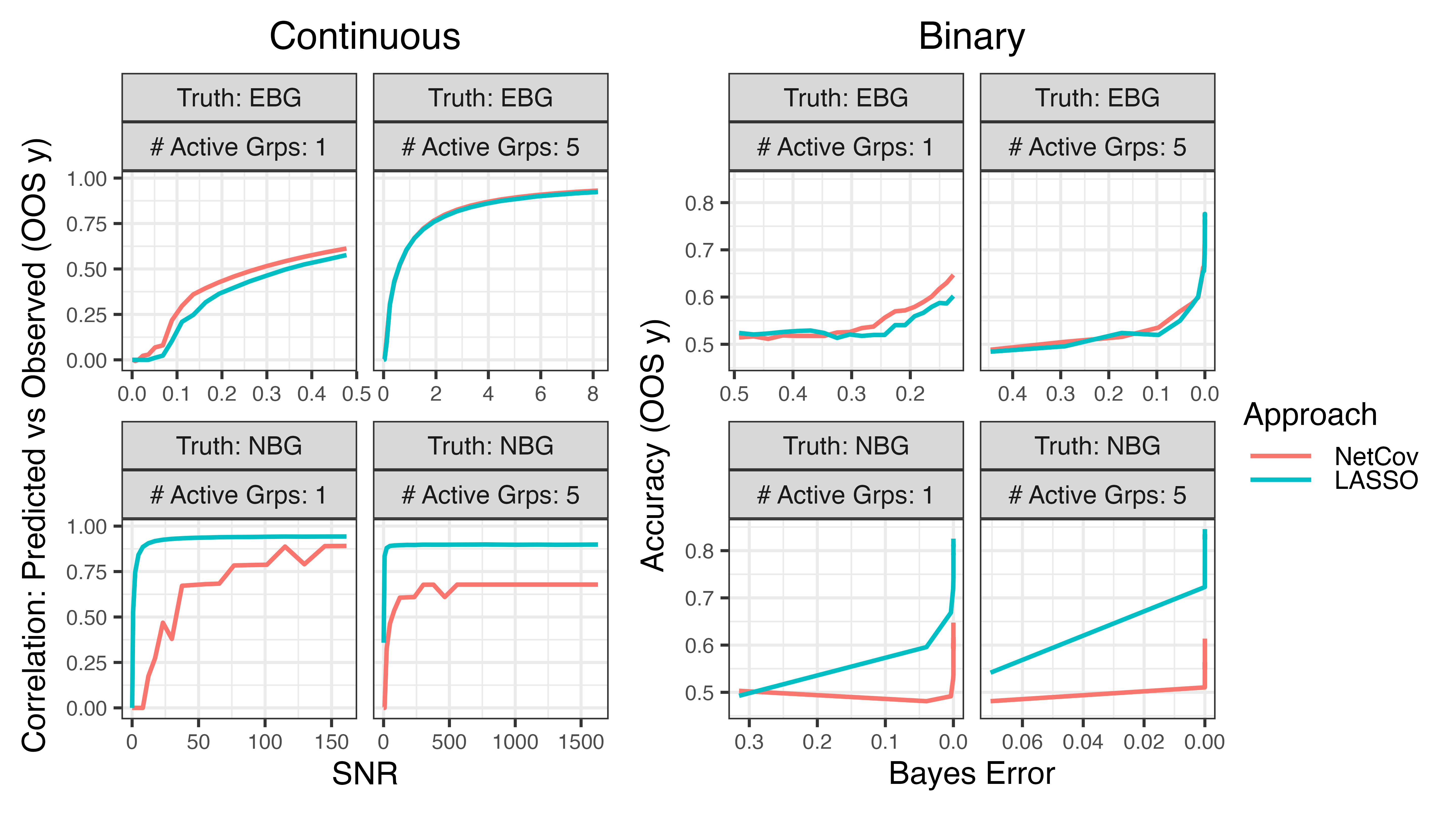}
  \caption{
    Out-of-sample prediction performance in Experiment II by NetCov (red) and LASSO (blue), as a function of problem difficulty (SNR for continuous response and Bayes error for binary).    Note the horizontal scale is different in each panel.
  }
  \label{fig:exp-III-pred}
\end{figure}

Support recovery is depicted in \cref{fig:exp-III-support}.
As we see, NetCov with EBG generally outperforms the LASSO,
especially for recall at low signal strength regimes.  %
NetCov's EBG variant generally outperforms or is comparable to the LASSO with respect to out-of-sample performance for both continuous and binary responses as seen in \cref{fig:exp-III-pred}.
NBG is limited, unless the sample size is very large relative to the number of features or we add another penalty to encourage within-group sparsity.

\section{Application to Neuroimaging Data}
\label{sec:netcov-application}

We demonstrate the utility of our method by applying it to a subset of data from the Human Connectome Project \citep[HCP;][]{vanessenWUMinnHumanConnectome2013} obtained and processed by the lab of our collaborator (see Acknowledgments).
In brief, each participant contributes an observation
$\left( \M{A}^{(i)}, \M{X}^{(i)}, y^{(i)} \right)$
which comprises functional connectivity from resting state data,
activation during a working memory task,
and a variety of behavioral measures, respectively.
There are \num{881} participants that have complete data for all the measures we consider.
We partition these into a training set of size \num{785} participants and a test set of size \num{96}.
This particular training/test split corresponds to the partitions used in \citet{sripadaBasicUnitsInterIndividual2019}, which was constructed to avoid any twins or sets of familially related individuals appearing in {\em both} the training and test sets.
We describe each component of the observations in more detail below.

For $\M{A}^{(i)}$ and $\M{X}^{(i)}$,
spatial locations of nodes as well as their community assignments were defined according to the ``Power parcellation'' \citep{powerFunctionalNetworkOrganization2011}, mentioned earlier in the text.
This yields \num{264} nodes assigned to \num{14} communities.
Since our grouping scheme assumes that the nodes in a given system are meaningfully related, we removed all nodes that were assigned the ``unknown'' community label.
This left \num{236} nodes divided into \num{13} communities. 
The putative brain systems corresponding to the communities, and the number of nodes in each, are given in \cref{tab:power-communities}.

The connectivity data is a subset of that used in \citet{sripadaBasicUnitsInterIndividual2019}, which describes the processing pipeline in detail.
In brief, resting state fMRI data was obtained for each participant in 4 different sessions (two back-to-back sessions per day across two days), from which connectivity measures $\M{A}^{(i)}$ are extracted.
During a resting state fMRI scanning session, the participant's brain activity is indirectly measured at many thousands of voxels in the brain while they lie passively in the scanner.
Each of the 236 nodes corresponds to a set of voxels.
As part of a comprehensive preprocessing pipeline, average time courses are extracted for all voxels in the same node.
Each entry of $\M{A^{(i)}}$ is taken to be the correlation between the average time series at two of these nodes, Fisher transformed to the real line, for the $i$-th participant.

We form $\M{X}^{(i)}$ by obtaining a single node covariate ($d = 1$) for each of the \num{236} nodes from brain activity during the ``N-back'' task \citep{barchFunctionHumanConnectome2013},
which is designed to measure working memory.
This data is a superset\footnotemark{} of that used in \citet{panigrahi2020inference}, which describes the data further.
\footnotetext{In their analysis, \citet{panigrahi2020inference} used only data from the \num{785} participants in our training set.}
During the ``N-back'' task, participants view a sequence of images that are presented in blocks.
Each block corresponds to a condition.
In the 0-back condition, participants are asked to judge whether each presented item is the same as what they saw at the beginning of the block.
During the 2-back condition, participants indicate whether each item is the same as what they saw two trials previous.
Of course, the 2-back condition is more demanding with respect to working memory.
In an attempt to isolate brain activity specific to working memory, activation during the 0-back condition is subtracted from activation during the 2-back condition.
This removes activity common to both conditions (e.g., visual processing, motor activity to push a button, etc.).
This 2-back minus 0-back contrast was computed by our collaborator using in-house processing scripts that use SPM12.
These contrasts were initially computed at the voxel level, but averaged values were extracted for each of the 236 nodes using the MarsBar utility \citep{marsbar}.

For $y^{(i)}$, we separately consider various phenotypes provided by our collaborator.
Several of the responses reflect performance during tasks from the NIH Toolbox \citep{hodes2013NIHToolboxSetting}, namely the
\begin{enumerate*}[(i), series=phenos]
\item pattern comparison processing speed test (Processing Speed)
\item flanker inhibitory control and attention test (Flanker)
\item list sorting working memory test (List Sorting),
\end{enumerate*}
as well as
\begin{enumerate*}[resume*=phenos]
\item performance on the Penn Progressive Matrices task (PMAT) \citep{pmat}.
\end{enumerate*}
Other responses capture the five facets of the NEO personality assessment \citep{MCCRAE2004587}:
\begin{enumerate*}[resume*=phenos]
\item openness to experience (NEO: O)
\item conscientiousness (NEO: C)
\item extraversion (NEO: E)
\item agreeableness (NEO: A)
\item neuroticism (NEO: N).
\end{enumerate*}
We also considered
\begin{enumerate*}[resume*=phenos]
\item accuracy on the ``N-back'' task described above (Working Memory).
\end{enumerate*}
Finally, we considered
\begin{enumerate*}[resume*=phenos]
\item a measure of general cognitive ability (GCA) obtained from factor analysis described in \citet{sripadaTreadmillTestCognition2020}.
\end{enumerate*}
All candidate responses are continuous, so we use a linear model for the response, i.e.,
\begin{align*}
  y^{(i)} &= \V{Z}^{(i)} \V{\beta} + \epsilon^{(i)}.
\end{align*}

In addition to covariates of interest, there are several nuisance covariates.
These are age (conventionally represented by a linear and a quadratic term), handedness, gender, brain size, which multiband reconstruction algorithm was used, and movement of the head during resting state scan (``meanFD'') along with its square.
We control for these by first fitting a regression model to the training data that includes only the nuisance covariates and predicts the brain features and phenotypes.
Using the coefficients learned in the training data, we then subtract the nuisance-predicted values from both the training and test data and use this corrected data for all downstream tasks.

While we can assess out-of-sample performance on the test set, we cannot assess support recovery directly as we did in our simulation studies, since the true $\V{\beta}$ is unknown.
As in the simulations, we assess performance by computing the correlation coefficient between predicted and observed responses on the test data, in order to facilitate comparisons with the neuroimaging literature, which frequently uses this measure.   

We compare our approach with both the conventional LASSO as well as connectome predictive modeling (CPM).
CPM is a popular and relatively simple technique for predicting scalar responses using brain connectivity data. In brief \citep[see][for a more detailed explanation]{shenUsingConnectomebasedPredictive2017}, it is a three-step procedure that involves marginal feature selection, feature aggregation, and then estimation of a regression model.
The first step consists of correlating screening between edge weights of the connectome and the response $y$.
Next, all edges that pass screening are aggregated into two summary measures, by summing together those that are positively correlated with the response, and those that are negatively correlated.
In the last step, a simple regression model is fit with these two summary measures as predictors.
While there are a number of variations (involving, e.g., robust regression), we opt for this simple pipeline and use $p < 0.01$ as the threshold for feature selection.
One noteworthy limitation of CPM, in contrast to NetCov, is that it operates only on edge weights and does not make use of node covariates.
For NetCov and LASSO, we use the same approach as in \cref{sec:netcov-numer-exper}, including cross-validation on the training data to select $\hat{\lambda}$.

\begin{figure}[ht]
  \centering
  \includegraphics[width=\textwidth]{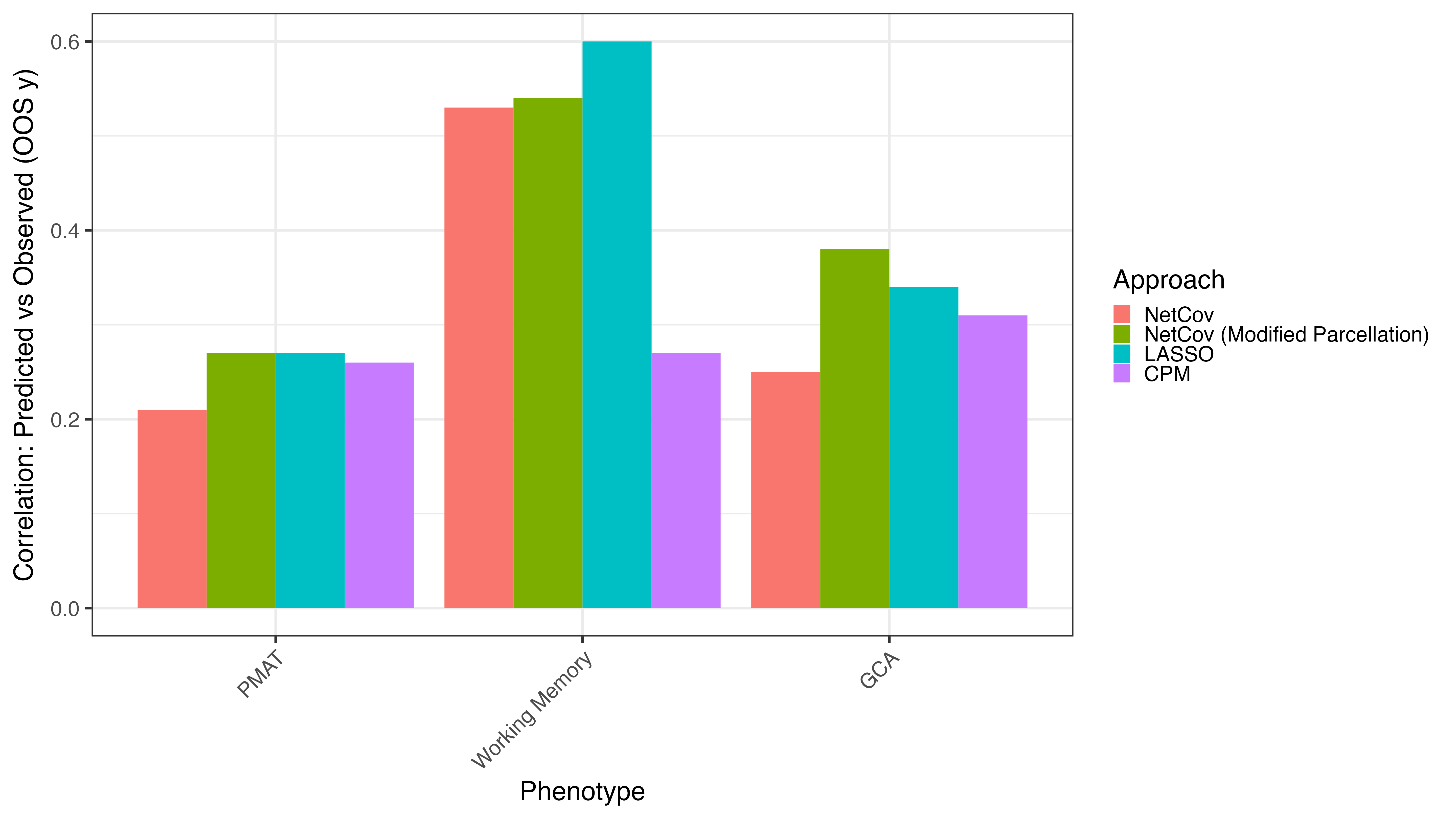}
  \caption{Out-of-sample correlation for selected phenotypes in application to human neuroimaging data.  }
  \label{fig:hcp-rho}
\end{figure}

We show results for selected phenotypes in \cref{fig:hcp-rho}.
Tuning of $\lambda$ yields poor performance for other phenotypes, and we present these results in Appendix~\ref{sec:netcov-hcp-results-app}.
For the original Power parcellation community assignments, NetCov is reasonably competitive with, although typically slightly worse than, the LASSO.
When we use the modified Power parcellation community assignments as described in \cref{sec:netcov-experiment-iii}, this small difference disappears.
Good predictive performance of the LASSO is expected since it imposes less structure on the estimated coefficients, at the cost of less interpretability.

\newcommand{\takcaption}[1]{
  Visualization of $\V{\beta}$ Coefficients with ``#1'' as Response.
  Coefficients from NetCov with EBG are presented at left ($\M{\beta}_{\M{X}}$) and on the lower triangle ($\M{\beta}_{\M{A}}$).
  Coefficients from LASSO are presented at right ($\M{\beta}_{\M{X}}$) and on the upper triangle ($\M{\beta}_{\M{A}}$).
}

\newcommand{\takmodcaption}{
  Faint, dashed lines depict boundaries of subcommunities for the modified Power parcellation (see \cref{sec:netcov-experiment-iii}).
}

\begin{figure}[ht]
  \centering
  \includegraphics[width=1.0\textwidth]{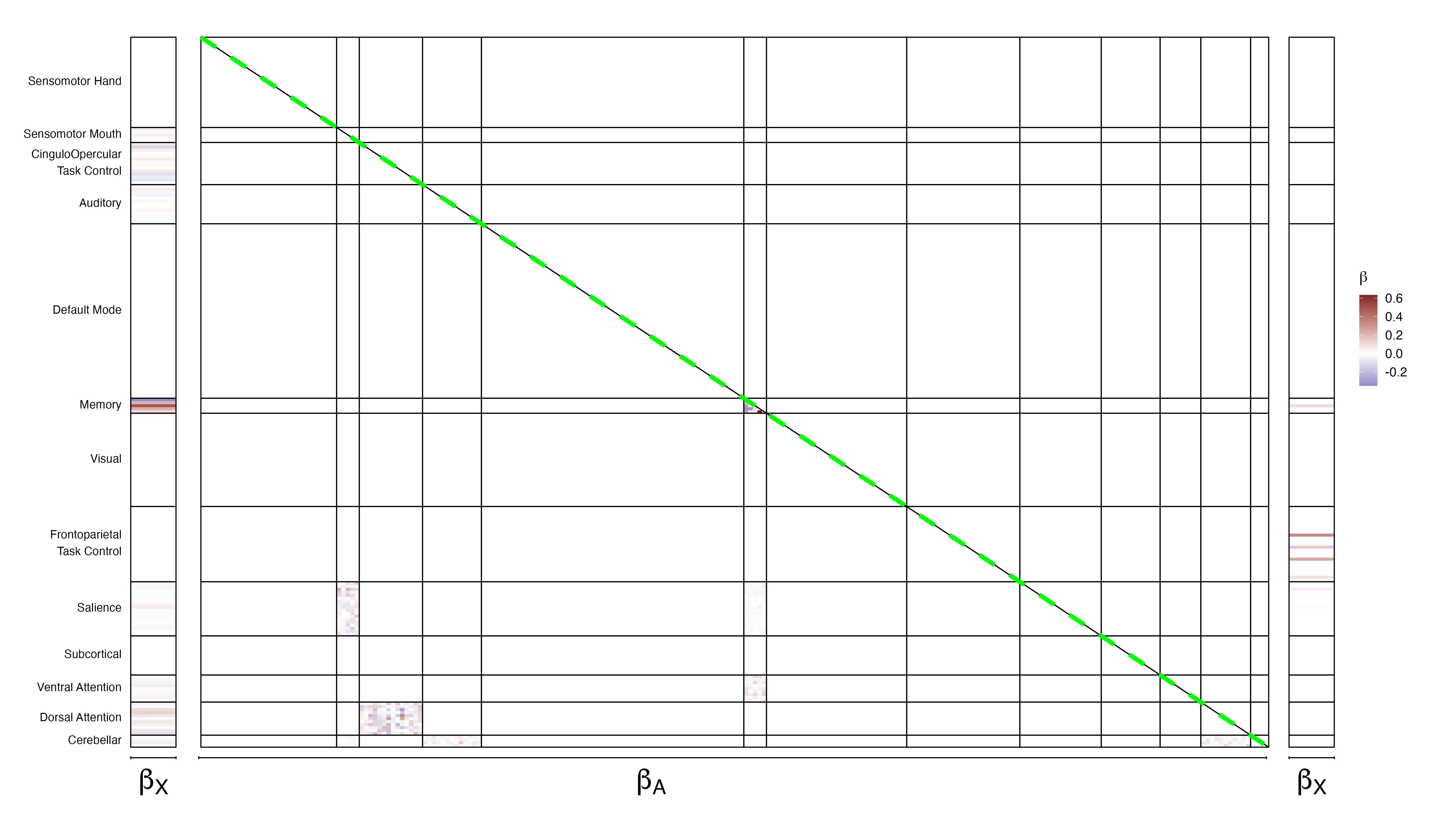}
  \caption{
    Visualization of $\V{\beta}$ coefficients with GCA as response.
    Coefficients from NetCov with EBG are presented at left ($\M{\beta}_{\M{X}}$) and on the lower triangle ($\M{\beta}_{\M{A}}$).
    Coefficients from LASSO are presented at right ($\M{\beta}_{\M{X}}$) and on the upper triangle ($\M{\beta}_{\M{A}}$).
    Solid lines depict boundaries of the Power parcellation.
  }
  \label{fig:hcp-coef-g}
\end{figure}

We present the estimated coefficients for $\M{\beta}_{\M{A}}$ and $\M{\beta}_{\M{X}}$ associated with GCA in \cref{fig:hcp-coef-g} and present results from CPM and for other phenotypes in \cref{sec:netcov-hcp-results-app}.
These figures illustrate the degree to which NetCov yields more interpretable solutions, implicating only a small number of brain systems through both edge weights and node covariates.
In contrast, non-zero LASSO coefficients are scattered across brain systems and moreover appear not to include any edges.
Of particular note, some of the systems that NetCov selects (e.g., Cingulo-Opercular Task Control and Dorsal Attention), are consistent with previous reports in the literature that studied the neural basis of general intelligence, closely related to our GCA factor \citep{duncan2000NeuralBasisGeneral,tong2022TransdiagnosticConnectomeSignatures}.

\section{Summary and Discussion}
\label{sec:netcov-summary-discussion}

We have introduced NetCov, a method for prediction from samples of weighted networks and node covariates, which offers a novel way to discover relevant brain systems when both edge and node covariates are present.
We proposed two approaches, node-based and edge-based (NBG and EBG), depending on what we believe is central to the mechanism of the underlying model,
and implemented both through constructing appropriate group penalties.  
As we saw in simulations, when the assumptions underlying these grouping schemes are met, NetCov yields both superior predictive performance and better support recovery.
In our application to human neuroimaging data, NetCov offers comparable predictive performance to the LASSO but dramatically enhances interpretability of findings.

There are some settings that present challenges for NetCov.
As demonstrated in \cref{sec:netcov-experiment-ii} where we carried out Experiment II but with the original community assignments,
large groups with more features than observations can be problematic,
especially for NBG,
and as the number of nodes $n$ grows, the size of NBG groups can grow as fast as $O(n^2)$.
EBG suffers less from this and this issue can be addressed by the use of smaller communities, but constructing them of course needs to be application-specific.
This challenge, faced by the group LASSO generally, is discussed in \citet[page 250, see citations within]{buhlmannStatisticsHighDimensionalData2011}, where one proposed remedy is to use the smoothed group LASSO instead.
An alternative approach which may preserve interpretability while overcoming these limitations, is the use of ``bi-level'' selection methods \citep{huangGroupBridgeApproach2009,brehenyPenalizedMethodsBilevel2009,brehenyGroupExponentialLasso2015} or the sparse group LASSO \citep{friedmanNoteGroupLasso2010,caiSparseGroupLasso2019}.
These methods select both feature groups and smaller subsets of features within each group;
examples of this approach include
\citet{relion2019NetworkClassificationApplications,richie-halfordMultidimensionalAnalysisDetection2021}.
Another alternative is the ridged group LASSO \citep{simonStandardizationGroupLasso2012},
which performs selection at the group level and scales the penalty applied to standardized groups on their ``effective degrees of freedom.''

Future work in the neuroimaging setting includes application of this method to other datasets, especially where the outcome is a binary indicator of a degenerative disease process, since this is one of the chief motivations for both the NBG and EBG schemes.
In addition, there are numerous alternative candidates for node covariates, including structural measures like gray matter volume or surface measures like cortical thickness.
Finally, there are also other candidate edge weights, including those obtained from diffusion weighted imaging (DWI) which is believed to capture anatomical, rather than functional, features of brain connectivity.
If multiple types of edge weights were present, both NBG and EBG could be extended to accommodate their simultaneous use.

While the selection of interpretable feature groups is a useful step on the path to a better understanding of complex phenomena, NetCov does not directly provide inference.
Although data splitting can be employed, wherein a subset of features is selected in training data, and then an appropriately restricted model is fit in test data with conventional inferential tests, recent results from post-selection inference for the group LASSO enable inferential tests at the level of feature groups \citep{yangSelectiveInferenceGroupsparse2016} as well as individual coefficients within feature groups \citep{panigrahi2020inference}.
These methods could be developed for NetCov to perform inference in addition to variable selection.  

\section*{Acknowledgments}
This research was supported by NSF grants DMS-1521551, DMS-1646108, DMS-1916222, and DMS-2052632 and a grant from the Dana Foundation.  D.~K.~was also supported by a Rackham Predoctoral Fellowship awarded by the University of Michigan. K.~L. was also supported by the University of Wisconsin--Madison Office of the Vice Chancellor for Research and Graduate Education with funding from the Wisconsin Alumni Research Foundation.
Data were provided in part by the Human Connectome Project, WU-Minn Consortium (Principal Investigators: David Van Essen and Kamil Ugurbil; 1U54MH091657) funded by the 16 NIH Institutes and Centers that support the NIH Blueprint for Neuroscience Research; and by the McDonnell Center for Systems Neuroscience at Washington University. %
We thank Dr.~Chandra Sripada and his research group (with particular thanks to Saige Rutherford) for providing us with a processed version of the data as well as helpful comments in interpreting our results.
This research was supported in part through computational resources and services provided by Advanced Research Computing (ARC), a division of Information and Technology Services (ITS) at the University of Michigan, Ann Arbor.

\bibliographystyle{apalike}

\bibliography{refs}

\begin{thebibliography}{}

\bibitem[Arbabshirani et~al., 2017]{ARBABSHIRANI2017137}
Arbabshirani, M.~R., Plis, S., Sui, J., and Calhoun, V.~D. (2017).
\newblock Single subject prediction of brain disorders in neuroimaging:
  Promises and pitfalls.
\newblock {\em NeuroImage}, 145:137--165.
\newblock Individual Subject Prediction.

\bibitem[Barch et~al., 2013]{barchFunctionHumanConnectome2013}
Barch, D.~M., Burgess, G.~C., Harms, M.~P., Petersen, S.~E., Schlaggar, B.~L.,
  Corbetta, M., Glasser, M.~F., Curtiss, S., Dixit, S., Feldt, C., Nolan, D.,
  Bryant, E., Hartley, T., Footer, O., Bjork, J.~M., Poldrack, R., Smith, S.,
  {Johansen-Berg}, H., Snyder, A.~Z., and Van~Essen, D.~C. (2013).
\newblock Function in the human connectome: {{Task}}-{{fMRI}} and individual
  differences in behavior.
\newblock {\em NeuroImage}, 80:169--189.

\bibitem[Bilker et~al., 2012]{pmat}
Bilker, W.~B., Hansen, J.~A., Brensinger, C.~M., Richard, J., Gur, R.~E., and
  Gur, R.~C. (2012).
\newblock Development of abbreviated nine-item forms of the raven’s standard
  progressive matrices test.
\newblock {\em Assessment}, 19(3):354--369.
\newblock PMID: 22605785.

\bibitem[Breheny, 2015]{brehenyGroupExponentialLasso2015}
Breheny, P. (2015).
\newblock The group exponential lasso for bi-level variable selection.
\newblock {\em Biometrics}, 71(3):731--740.

\bibitem[Breheny and Huang, 2009]{brehenyPenalizedMethodsBilevel2009}
Breheny, P. and Huang, J. (2009).
\newblock Penalized methods for bi-level variable selection.
\newblock {\em Statistics and Its Interface}, 2(3):369--380.

\bibitem[Breheny and Huang, 2015]{brehenyGroupDescentAlgorithms2015}
Breheny, P. and Huang, J. (2015).
\newblock Group descent algorithms for nonconvex penalized linear and logistic
  regression models with grouped predictors.
\newblock {\em Statistics and Computing}, 25(2):173--187.

\bibitem[Brett et~al., 2002]{marsbar}
Brett, M., Anton, J.-L., Valabregue, R., and Poline, J.-B. (2002).
\newblock Region of interest analysis using an {SPM} toolbox.
\newblock In {\em Presented at the 8th International Conference on Functional
  Mapping of the Human Brain}.
\newblock Abstract Available in NeuroImage, Vol 16, No 2.

\bibitem[Burgos and Colliot, 2020]{burgosMachineLearningClassification2020}
Burgos, N. and Colliot, O. (2020).
\newblock Machine learning for classification and prediction of brain diseases:
  Recent advances and upcoming challenges.
\newblock {\em Current Opinion in Neurology}, 33(4):439--450.

\bibitem[Bühlmann and van~de Geer,
  2011]{buhlmannStatisticsHighDimensionalData2011}
Bühlmann, P. and van~de Geer, S. (2011).
\newblock {\em Statistics for {{High}}-{{Dimensional Data}}}.
\newblock Springer {{Series}} in {{Statistics}}. Springer Berlin Heidelberg.

\bibitem[Cai et~al., 2022]{caiSparseGroupLasso2019}
Cai, T.~T., Zhang, A., and Zhou, Y. (2022).
\newblock Sparse {{Group Lasso}}: {{Optimal Sample Complexity}}, {{Convergence
  Rate}}, and {{Statistical Inference}}.
\newblock {\em {IEEE} Transactions on Information Theory}, 68(9):5975--6002.

\bibitem[Calhoun et~al., 2017]{CALHOUN2017135}
Calhoun, V.~D., Lawrie, S.~M., Mourao-Miranda, J., and Stephan, K.~E. (2017).
\newblock Prediction of individual differences from neuroimaging data.
\newblock {\em NeuroImage}, 145:135--136.
\newblock Individual Subject Prediction.

\bibitem[Calhoun and Sui, 2016]{calhounMultimodalFusionBrain2016}
Calhoun, V.~D. and Sui, J. (2016).
\newblock Multimodal {{Fusion}} of {{Brain Imaging Data}}: {{A Key}} to
  {{Finding}} the {{Missing Link}}(s) in {{Complex Mental Illness}}.
\newblock {\em Biological Psychiatry: Cognitive Neuroscience and Neuroimaging},
  1(3):230--244.

\bibitem[Chung et~al., 2021]{chungStatisticalConnectomics2021}
Chung, J., Bridgeford, E., Arroyo, J., Pedigo, B.~D., {Saad-Eldin}, A.,
  Gopalakrishnan, V., Xiang, L., Priebe, C.~E., and Vogelstein, J.~T. (2021).
\newblock Statistical {{Connectomics}}.
\newblock {\em Annual Review of Statistics and Its Application}, 8(1):463--492.

\bibitem[Duncan et~al., 2000]{duncan2000NeuralBasisGeneral}
Duncan, J., Seitz, R.~J., Kolodny, J., Bor, D., Herzog, H., Ahmed, A., Newell,
  F.~N., and Emslie, H. (2000).
\newblock A {Neural} {Basis} for {General} {Intelligence}.
\newblock {\em Science}, 289(5478):457--460.
\newblock Publisher: American Association for the Advancement of Science.

\bibitem[Fox, 2018]{foxMappingSymptomsBrain2018}
Fox, M.~D. (2018).
\newblock Mapping {{Symptoms}} to {{Brain Networks}} with the {{Human
  Connectome}}.
\newblock {\em New England Journal of Medicine}, 379(23):2237--2245.

\bibitem[Friedman et~al., 2010a]{friedmanNoteGroupLasso2010}
Friedman, J., Hastie, T., and Tibshirani, R. (2010a).
\newblock A note on the group lasso and a sparse group lasso.
\newblock {\em arXiv:1001.0736}.

\bibitem[Friedman et~al., 2010b]{friedmanRegularizationPathsGeneralized2010}
Friedman, J., Hastie, T., and Tibshirani, R. (2010b).
\newblock Regularization {{Paths}} for {{Generalized Linear Models}} via
  {{Coordinate Descent}}.
\newblock {\em Journal of Statistical Software}, 33(1):1--22.

\bibitem[Ginestet et~al., 2017]{ginestetHypothesisTestingNetwork2017}
Ginestet, C.~E., Li, J., Balachandran, P., Rosenberg, S., and Kolaczyk, E.~D.
  (2017).
\newblock Hypothesis testing for network data in functional neuroimaging.
\newblock {\em The Annals of Applied Statistics}, 11(2):725--750.

\bibitem[Hodes et~al., 2013]{hodes2013NIHToolboxSetting}
Hodes, R.~J., Insel, T.~R., Landis, S.~C., and Research, O. b. o. t. N. B.
  f.~N. (2013).
\newblock The {{NIH Toolbox}}: {{Setting}} a standard for biomedical research.
\newblock {\em Neurology}, 80(11 Supplement 3):S1--S1.

\bibitem[Hsu et~al., 2018]{hsuRestingstateFunctionalConnectivity2018}
Hsu, W.-T., Rosenberg, M.~D., Scheinost, D., Constable, R.~T., and Chun, M.~M.
  (2018).
\newblock Resting-state functional connectivity predicts neuroticism and
  extraversion in novel individuals.
\newblock {\em Social Cognitive and Affective Neuroscience}, 13(2):224--232.

\bibitem[Huang et~al., 2009]{huangGroupBridgeApproach2009}
Huang, J., Ma, S., Xie, H., and Zhang, C.-H. (2009).
\newblock A group bridge approach for variable selection.
\newblock {\em Biometrika}, 96(2):339--355.

\bibitem[Huang and Zhang, 2010]{huangBenefitGroupSparsity2010}
Huang, J. and Zhang, T. (2010).
\newblock The benefit of group sparsity.
\newblock {\em The Annals of Statistics}, 38(4):1978--2004.

\bibitem[Jacob et~al., 2009]{jacobGroupLassoOverlap2009}
Jacob, L., Obozinski, G., and Vert, J.-P. (2009).
\newblock Group {{Lasso}} with {{Overlap}} and {{Graph Lasso}}.
\newblock In {\em Proceedings of the 26th {{Annual International Conference}}
  on {{Machine Learning}}}, {{ICML}} '09, pages 433--440. ACM.

\bibitem[Khosla et~al., 2019]{khoslaMachineLearningRestingstate2019}
Khosla, M., Jamison, K., Ngo, G.~H., Kuceyeski, A., and Sabuncu, M.~R. (2019).
\newblock Machine learning in resting-state {{fMRI}} analysis.
\newblock {\em Magnetic Resonance Imaging}, 64:101--121.

\bibitem[Kim et~al., 2023]{kimGraphawareModelingBrain2021}
Kim, Y., Kessler, D., and Levina, E. (2023).
\newblock Graph-aware {{Modeling}} of {{Brain Connectivity Networks}}.
\newblock {\em The Annals of Applied Statistics}.

\bibitem[Lang et~al., 2017]{langBatchtoolsToolsWork2017}
Lang, M., Bischl, B., and Surmann, D. (2017).
\newblock Batchtools: {{Tools}} for {{R}} to work on batch systems.
\newblock {\em Journal of Open Source Software}, 2(10):135.

\bibitem[McCrae and Costa, 2004]{MCCRAE2004587}
McCrae, R.~R. and Costa, P.~T. (2004).
\newblock A contemplated revision of the neo five-factor inventory.
\newblock {\em Personality and Individual Differences}, 36(3):587--596.

\bibitem[McCullagh and Nelder, 1998]{mccullaghGeneralizedLinearModels1998}
McCullagh, P. and Nelder, J.~A. (1998).
\newblock {\em Generalized Linear Models}.
\newblock Number~37 in Monographs on Statistics and Applied Probability.
  Chapman \& Hall/CRC, 2nd ed edition.

\bibitem[Nardi and Rinaldo, 2008]{nardiAsymptoticPropertiesGroup2008}
Nardi, Y. and Rinaldo, A. (2008).
\newblock On the asymptotic properties of the group lasso estimator for linear
  models.
\newblock {\em Electronic Journal of Statistics}, 2:605--633.

\bibitem[Noble et~al., 2022]{nobleImprovingPowerFunctional2022}
Noble, S., Mejia, A.~F., Zalesky, A., and Scheinost, D. (2022).
\newblock Improving power in functional magnetic resonance imaging by moving
  beyond cluster-level inference.
\newblock {\em Proceedings of the National Academy of Sciences},
  119(32):e2203020119.

\bibitem[Obozinski et~al., 2011]{obozinskiGroupLassoOverlaps2011}
Obozinski, G., Jacob, L., and Vert, J.-P. (2011).
\newblock Group {{Lasso}} with {{Overlaps}}: The {{Latent Group Lasso}}
  approach.
\newblock {\em arXiv:1110.0413}.

\bibitem[Panigrahi et~al., 2023]{panigrahi2020inference}
Panigrahi, S., MacDonald, P.~W., and Kessler, D. (2023).
\newblock Approximate post-selective inference for regression with the group
  lasso.
\newblock {\em Journal of Machine Learning Research}, 24(79):1--49.

\bibitem[Power et~al., 2011]{powerFunctionalNetworkOrganization2011}
Power, J.~D., Cohen, A.~L., Nelson, S.~M., Wig, G.~S., Barnes, K.~A., Church,
  J.~A., Vogel, A.~C., Laumann, T.~O., Miezin, F.~M., Schlaggar, B.~L., and
  Petersen, S.~E. (2011).
\newblock Functional {{Network Organization}} of the {{Human Brain}}.
\newblock {\em Neuron}, 72(4):665--678.

\bibitem[{R Core Team}, 2023]{rcoreteamLanguageEnvironmentStatistical}
{R Core Team} (2023).
\newblock {\em R: A Language and Environment for Statistical Computing}.
\newblock R Foundation for Statistical Computing, Vienna, Austria.

\bibitem[Reli{\'o}n et~al., 2019]{relion2019NetworkClassificationApplications}
Reli{\'o}n, J. D.~A., Kessler, D., Levina, E., and Taylor, S.~F. (2019).
\newblock Network classification with applications to brain connectomics.
\newblock {\em The Annals of Applied Statistics}, 13(3):1648--1677.

\bibitem[{Richie-Halford} et~al.,
  2021]{richie-halfordMultidimensionalAnalysisDetection2021}
{Richie-Halford}, A., Yeatman, J.~D., Simon, N., and Rokem, A. (2021).
\newblock Multidimensional analysis and detection of informative features in
  human brain white matter.
\newblock {\em PLOS Computational Biology}, 17(6):e1009136.

\bibitem[Shen et~al., 2017]{shenUsingConnectomebasedPredictive2017}
Shen, X., Finn, E.~S., Scheinost, D., Rosenberg, M.~D., Chun, M.~M.,
  Papademetris, X., and Constable, R.~T. (2017).
\newblock Using connectome-based predictive modeling to predict individual
  behavior from brain connectivity.
\newblock {\em Nature Protocols}, 12(3):506--518.

\bibitem[Shimizu et~al., 2015]{shimizuProbabilisticDiagnosisUnderstanding2015a}
Shimizu, Y., Yoshimoto, J., Toki, S., Takamura, M., Yoshimura, S., Okamoto, Y.,
  Yamawaki, S., and Doya, K. (2015).
\newblock Toward {{Probabilistic Diagnosis}} and {{Understanding}} of
  {{Depression Based}} on {{Functional MRI Data Analysis}} with {{Logistic
  Group LASSO}}.
\newblock {\em PLOS ONE}, 10(5):e0123524.

\bibitem[Simon and Tibshirani, 2012]{simonStandardizationGroupLasso2012}
Simon, N. and Tibshirani, R. (2012).
\newblock Standardization and the {{Group Lasso Penalty}}.
\newblock {\em Statistica Sinica}, 22(3):983--1001.

\bibitem[Sripada et~al., 2019]{sripadaBasicUnitsInterIndividual2019}
Sripada, C., Angstadt, M., Rutherford, S., Kessler, D., Kim, Y., Yee, M., and
  Levina, E. (2019).
\newblock Basic {{Units}} of {{Inter}}-{{Individual Variation}} in {{Resting
  State Connectomes}}.
\newblock {\em Scientific Reports}, 9(1):1900.

\bibitem[Sripada et~al., 2020]{sripadaTreadmillTestCognition2020}
Sripada, C., Angstadt, M., Rutherford, S., Taxali, A., and Shedden, K. (2020).
\newblock Toward a ``treadmill test'' for cognition: {{Improved}} prediction of
  general cognitive ability from the task activated brain.
\newblock {\em Human Brain Mapping}, 41(12):3186--3197.

\bibitem[Tang et~al., 2017]{tangSemiparametricTwoSampleHypothesis2017}
Tang, M., Athreya, A., Sussman, D.~L., Lyzinski, V., Park, Y., and Priebe,
  C.~E. (2017).
\newblock A {{Semiparametric Two-Sample Hypothesis Testing Problem}} for
  {{Random Graphs}}.
\newblock {\em Journal of Computational and Graphical Statistics},
  26(2):344--354.

\bibitem[Tibshirani, 1996]{tibshiraniRegressionShrinkageSelection1996}
Tibshirani, R. (1996).
\newblock Regression {{Shrinkage}} and {{Selection}} via the {{Lasso}}.
\newblock {\em Journal of the Royal Statistical Society. Series B
  (Methodological)}, 58(1):267--288.

\bibitem[Tong et~al., 2022]{tong2022TransdiagnosticConnectomeSignatures}
Tong, X., Xie, H., Carlisle, N., Fonzo, G.~A., Oathes, D.~J., Jiang, J., and
  Zhang, Y. (2022).
\newblock Transdiagnostic connectome signatures from resting-state {fMRI}
  predict individual-level intellectual capacity.
\newblock {\em Translational Psychiatry}, 12(1):1--11.
\newblock Number: 1 Publisher: Nature Publishing Group.

\bibitem[Van~Essen et~al., 2013]{vanessenWUMinnHumanConnectome2013}
Van~Essen, D.~C., Smith, S.~M., Barch, D.~M., Behrens, T. E.~J., Yacoub, E.,
  Ugurbil, K., and {the WU-Minn HCP Consortium} (2013).
\newblock The {{WU}}-{{Minn Human Connectome Project}}: {{An}} overview.
\newblock {\em NeuroImage}, 80:62--79.

\bibitem[Wang et~al., 2022]{wangPriceCompetitionEffect2020}
Wang, H., Yang, Y., and Su, W.~J. (2022).
\newblock The {{Price}} of {{Competition}}: {{Effect Size Heterogeneity
  Matters}} in {{High Dimensions}}.
\newblock {\em {IEEE} Transactions on Information Theory}, 68(8):5268--5294.

\bibitem[Xia et~al., 2020]{xiaMultiscaleNetworkRegression2020}
Xia, C.~H., Ma, Z., Cui, Z., Bzdok, D., Thirion, B., Bassett, D.~S.,
  Satterthwaite, T.~D., Shinohara, R.~T., and Witten, D.~M. (2020).
\newblock Multi-scale network regression for brain-phenotype associations.
\newblock {\em Human Brain Mapping}.

\bibitem[Yang et~al., 2016]{yangSelectiveInferenceGroupsparse2016}
Yang, F., Foygel~Barber, R., Jain, P., and Lafferty, J. (2016).
\newblock Selective inference for group-sparse linear models.
\newblock In {\em Advances in {{Neural Information Processing Systems}}},
  volume~29. Curran Associates, Inc.

\bibitem[Yeo et~al., 2011]{yeoOrganizationHumanCerebral2011}
Yeo, B. T.~T., Krienen, F.~M., Sepulcre, J., Sabuncu, M.~R., Lashkari, D.,
  Hollinshead, M., Roffman, J.~L., Smoller, J.~W., Zöllei, L., Polimeni,
  J.~R., Fischl, B., Liu, H., and Buckner, R.~L. (2011).
\newblock The organization of the human cerebral cortex estimated by intrinsic
  functional connectivity.
\newblock {\em Journal of Neurophysiology}, 106(3):1125--1165.

\bibitem[Yu et~al., 2019]{yuChildhoodTraumaHistory2019}
Yu, M., Linn, K.~A., Shinohara, R.~T., Oathes, D.~J., Cook, P.~A., Duprat, R.,
  Moore, T.~M., Oquendo, M.~A., Phillips, M.~L., McInnis, M., Fava, M.,
  Trivedi, M.~H., McGrath, P., Parsey, R., Weissman, M.~M., and Sheline, Y.~I.
  (2019).
\newblock Childhood trauma history is linked to abnormal brain connectivity in
  major depression.
\newblock {\em Proceedings of the National Academy of Sciences},
  116(17):8582--8590.

\bibitem[Yuan and Lin, 2006]{yuan2006ModelSelectionEstimation}
Yuan, M. and Lin, Y. (2006).
\newblock Model selection and estimation in regression with grouped variables.
\newblock {\em Journal of the Royal Statistical Society: Series B (Statistical
  Methodology)}, 68(1):49--67.

\bibitem[Zeng and Breheny, 2016]{zengOverlappingGroupLogistic2016}
Zeng, Y. and Breheny, P. (2016).
\newblock Overlapping {{Group Logistic Regression}} with {{Applications}} to
  {{Genetic Pathway Selection}}.
\newblock {\em Cancer Informatics}, 15.

\end{thebibliography}

\appendix
\renewcommand{\figurename}{Supplementary Figure}
\renewcommand{\tablename}{Supplementary Table}
\section{Experimental Results for Semi-Synthetic Data with Original Community Membership}
\label{sec:netcov-experiment-ii}

\begin{table}[ht]
  \centering
  \begin{tabular}{r|l}
    Brain System Name & Number of Nodes \\
    \hline
    Sensomotor Hand & 30 \\
    Sensomotor Mouth & 5 \\
    Cingulo-Opercular Task Control & 14 \\
    Auditory & 13 \\
    Default Mode & 58 \\
    Memory & 5 \\
    Visual & 31 \\
    Frontoparietal Task Control & 25 \\
    Salience & 18 \\
    Subcortical & 13 \\
    Ventral Attention & 9 \\
    Dorsal Attention & 11 \\
    Cerebellar & 4
  \end{tabular}
  \caption{Systems (communities) in the Power parcellation and number of nodes in each \citep{powerFunctionalNetworkOrganization2011}.}
  \label{tab:power-communities}
\end{table}

Below, we present the results of an experiment analogous to that described in Section \ref{sec:netcov-experiment-iii} but where the \emph{original} community assignments as prescribed by the parcellation of \citet{powerFunctionalNetworkOrganization2011} are used.
As briefly mentioned in the main text,
this community assignment scheme yields groups with more predictors than observations.
For NBG, this is true for {\em all} groups---the NBG group based on the smallest community, with 4 nodes, has $\binom{4}{2} + 4 \times 232 + 4 = 938$ predictors, and $N_{\text{train}} = 785$---and the orthonormalization discussed in \cref{sec:netcov-stand-group-lasso} results in all of the feature groups being functionally identical.
For EBG, there are \num{11} groups with more predictors than observations, and so selecting any of these is problematic.
The covariance structure within the groups is nontrivial, and this is problematic for some of our larger groups because of the way the penalty \cref{eq:g-lasso} accounts for group sizes.

\begin{figure}[ht]
  \centering
  \includegraphics[width=\textwidth]{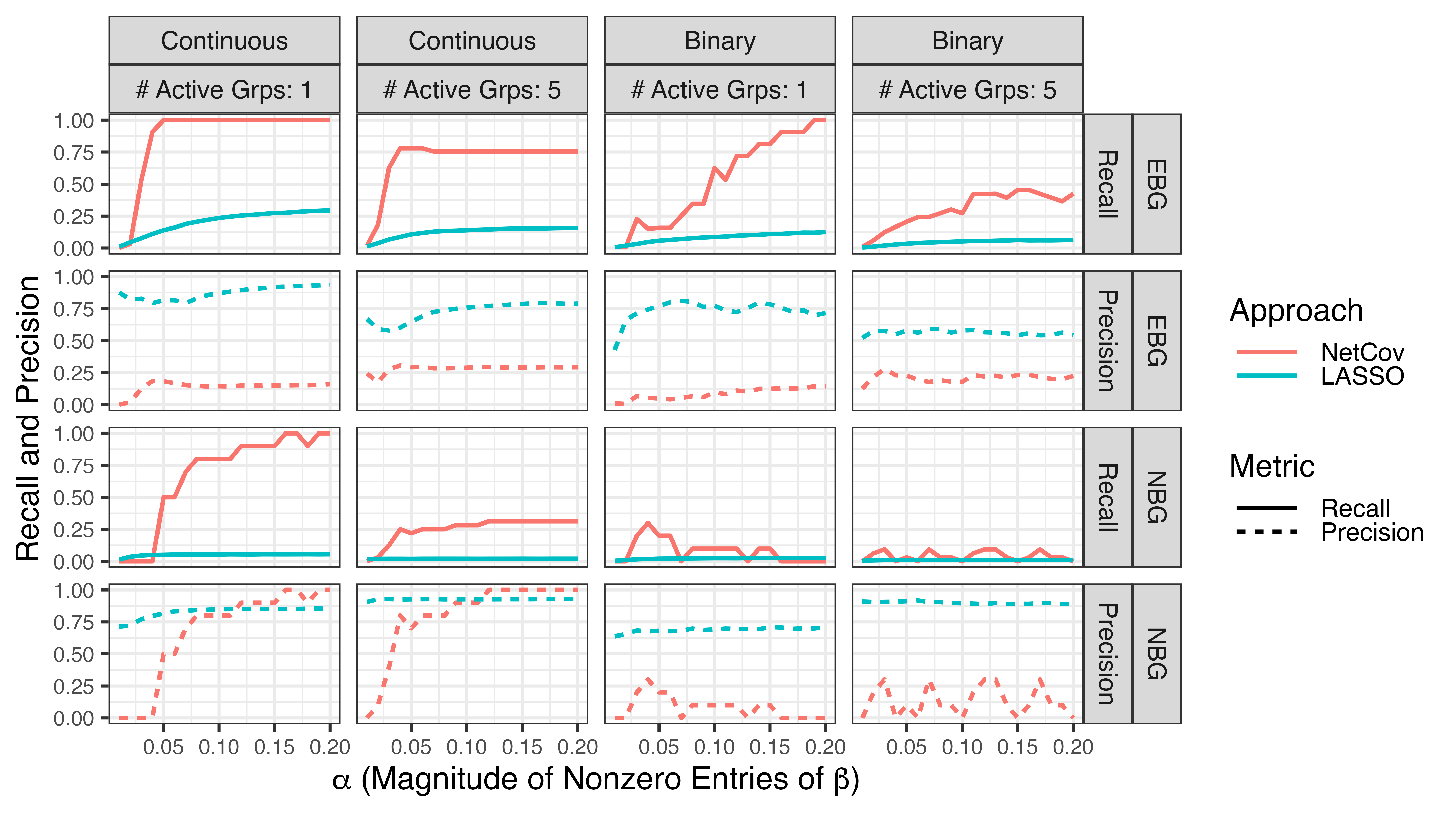}
  \caption{Support recovery in Experiment II with original communities: recall and precision as a function of nonzero coefficient magnitude $\alpha$ for NetCov (red) and LASSO (blue).  Each of the four columns corresponds to either continuous or binary response and either 1 or 5 active groups.  Each of the four rows corresponds to either EBG or NBG and either support recovery or precision for $\V{\beta}$.
  }
  \label{fig:exp-II-support}
\end{figure}

\begin{figure}[ht]
  \centering
  \includegraphics[width=\textwidth]{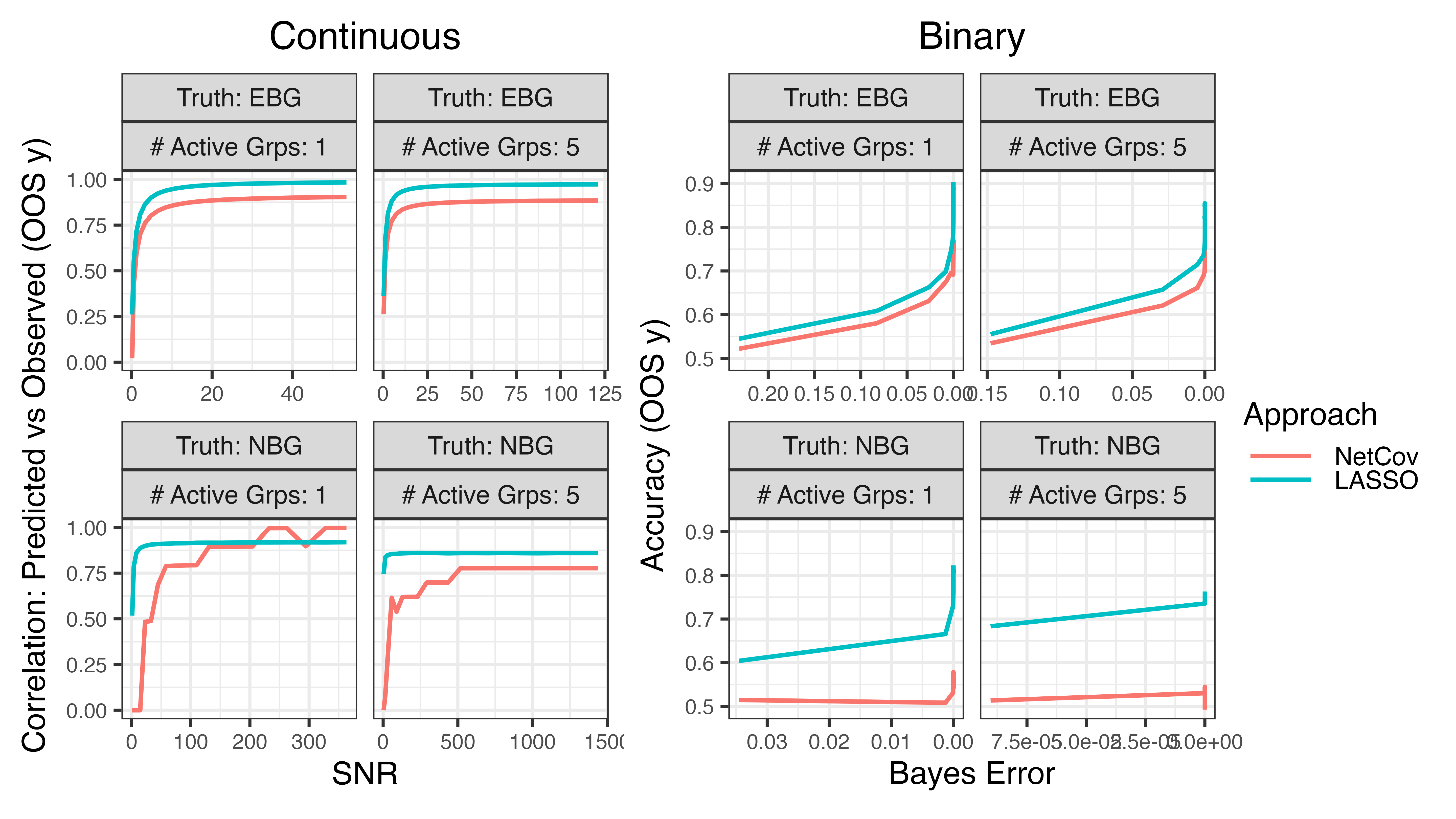}
  \caption{
    Out-of-sample prediction performance in Experiment II with original communities of NetCov (red) and LASSO (blue), as a function of problem difficulty (SNR for continuous response; Bayes error for binary).  Note that the horizontal scale is different in every panel.
  }
  \label{fig:exp-II-pred}
\end{figure}

In \cref{fig:exp-II-support} we see the results for support recovery of $\V{\beta}$.
NetCov does not show a consistent improvement over LASSO, and while at least for EBG, recall increases appreciably with growing signal strength, precision remains poor.
Performance of NBG is strikingly poor.
This pattern is present in both the continuous and binary response cases.
We believe that this is due to the presence of very large groups (especially for NBG), which are in turn due to the cardinality of communities as presented in \cref{tab:power-communities}.
The failure of NetCov to perform accurate support recovery limits its competitiveness for prediction in both the continuous and binary cases is depicted in \cref{fig:exp-II-pred}.
To overcome these challenges, we modified our parcellation to avoid the problems described above when we conducted Experiment II, described in the main text in Section
\cref{sec:netcov-experiment-iii}.

\section{ROC Curves}
\label{sec:netcov-roc-curves}
\begin{figure}[ht]
  \centering
  \includegraphics[width=0.8\textwidth]{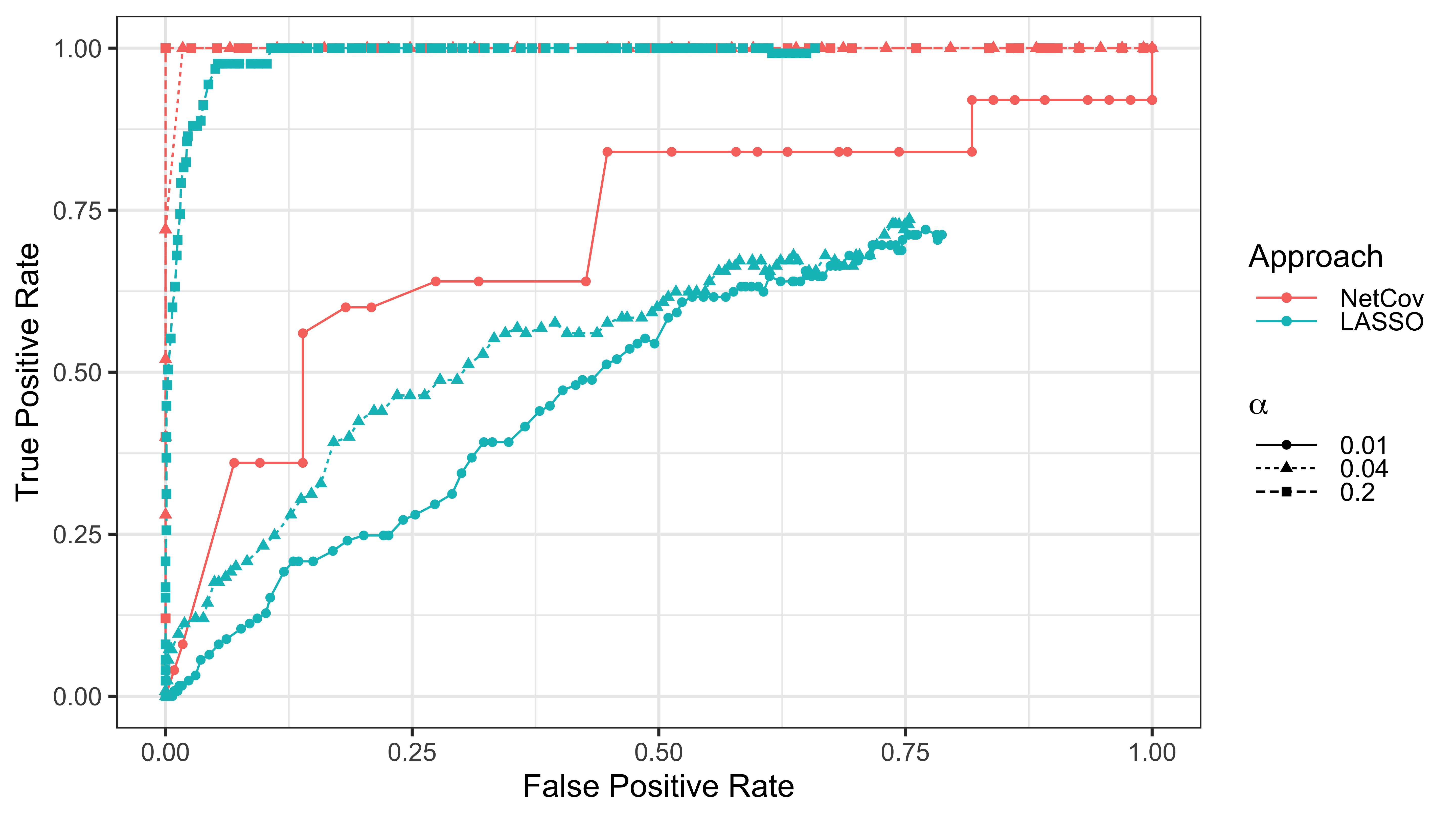}
  \caption{
    False positive and true positive rates along the $\lambda$ path.
    We depict receiver operating characteristic curves for the LASSO and NetCov: EBG model at varying levels of signal intensity.
    Data is drawn according to the setting of Experiment I as described in Section \ref{sec:netcov-experiment-i} with the EBG grouping scheme and \num{5} active groups.
  }
  \label{fig:roc}
\end{figure}

There is what might at first seem to be a surprising phenomenon present in our numerical experiments most visible in the ``\num{5} active groups'' panels of \cref{fig:exp-I-support}: as the magnitude of non-zero entries of $\beta$ grows, recall climbs steadily toward \num{1}, but precision climbs and then falls.
As discussed in the text, this behavior is a consequence of parameter tuning which selects the value of $\lambda$ that minimizes prediction risk rather than support recovery.
To further understand this, we conduct a small simulation study where we explore the behavior of the EBG variant of NetCov and the LASSO in this setting.
We obtain three realizations of data where $\beta$ follows the EBG grouping scheme, there are \num{5} active groups, and where we set $\alpha$ (the magnitude of active entries) to \num{.01}, \num{.04}, and \num{.2}, respectively; all other simulation parameters were fixed per the regime of Experiment I as described in \cref{sec:netcov-experiment-i}.
We fit both the LASSO and the NetCov method with the EBG grouping scheme to each of these realizations along a $\lambda$ path as described in \cref{sec:netcov-parameter-tuning} (note that the values of $\lambda$ along these paths for NetCov: EBG and LASSO will be distinct).
For a fixed value $\tilde{\lambda}$ along this path, we can obtain an estimated active set for both the LASSO and NetCov: EBG.
By comparing the estimated active set to the true active set, we compute both the True Positive Rate (True Positives divided by the sum of True and False Positives) and the False Positive Rate (False Positives divided by the sum of True and False Positives).
We connect these points to obtain six receiver operating characteristic curves for the three different levels of signal strength times the two different models.
This is depicted in \cref{fig:roc}.
Fixing $\alpha$ and comparing the two fitting methods, we see that the curve corresponding to NetCov: EBG is generally above or equal to the LASSO curve.
The apparently poor precision of the NetCov: EBG method in the high SNR regime (as may be especially evidence with \num{5} active groups as seen in \cref{fig:exp-I-support}) may seem odd given its apparently ``perfect'' ROC curve, but cross validation chooses $\hat{\lambda}$ somewhere along the curve where the True Positive Rate is \num{1}, but where the False Positive Rate is nonzero.
In other words, there exists a value of $\lambda$ that would give perfect support recovery, but cross validation does not choose it.
This appears to be a symptom of tuning $\lambda$ to minimize prediction error, and this leads to a too-small value of $\hat{\lambda}$ in order to avoid bias encountered from shrinking large predictors but this comes at the cost of selecting several inactive predictors.

\section{Additional Neuroimaging Results}
\label{sec:netcov-hcp-results-app}

\begin{figure}[ht]
  \centering
  \includegraphics[width=\textwidth]{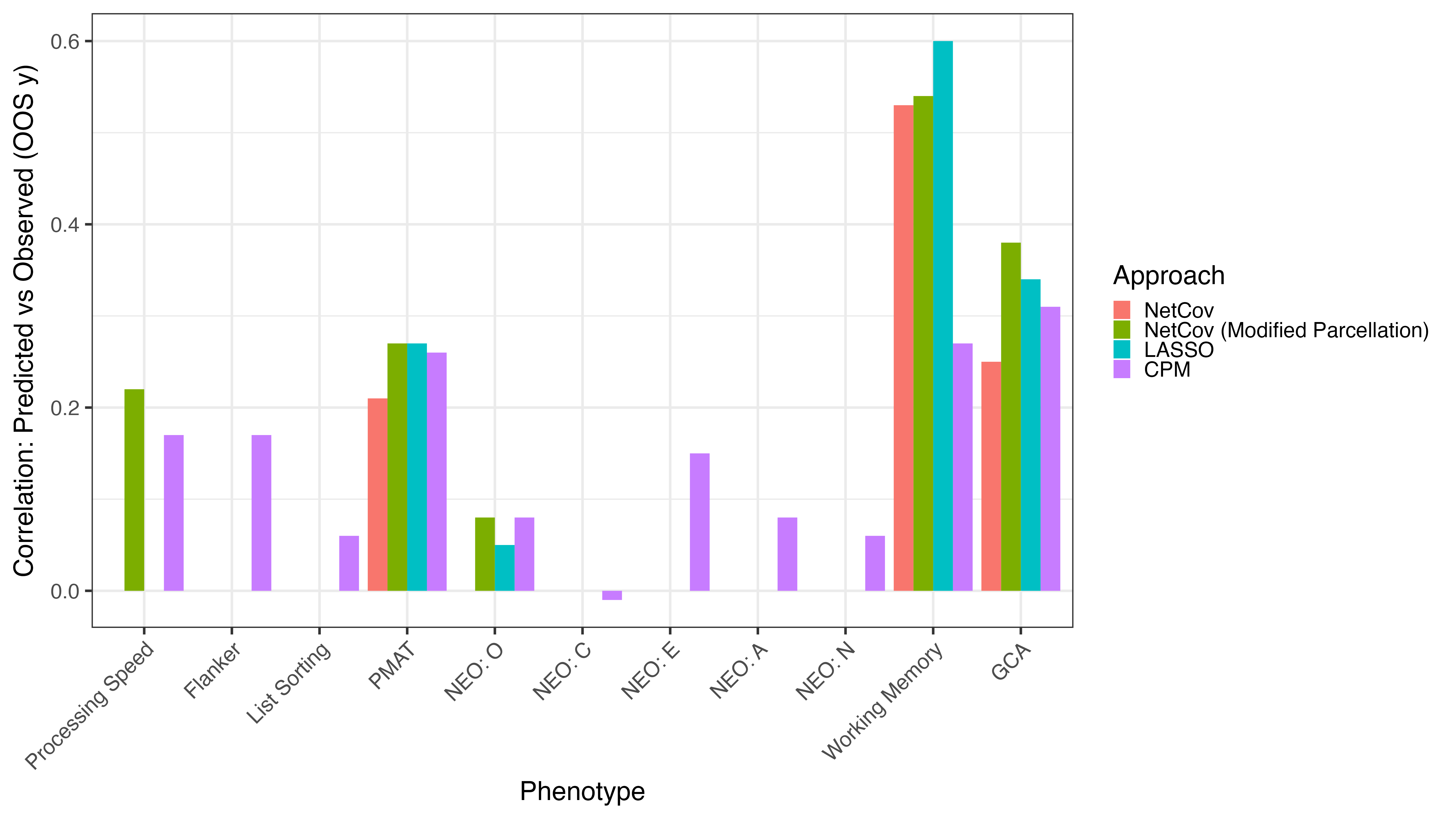}
  \caption{Out-of-sample correlation for all phenotypes in application to human neuroimaging data.}
  \label{fig:hcp-rho-all}
\end{figure}

In the main text, for brevity we presented out-of-sample correlation for a subset of the phenotypes in \cref{fig:hcp-rho}.
In \cref{fig:hcp-rho-all}, we present the out-of-sample correlations for all phenotypes.

\begin{figure}[ht]
  \centering
  \includegraphics[width=1.0\textwidth]{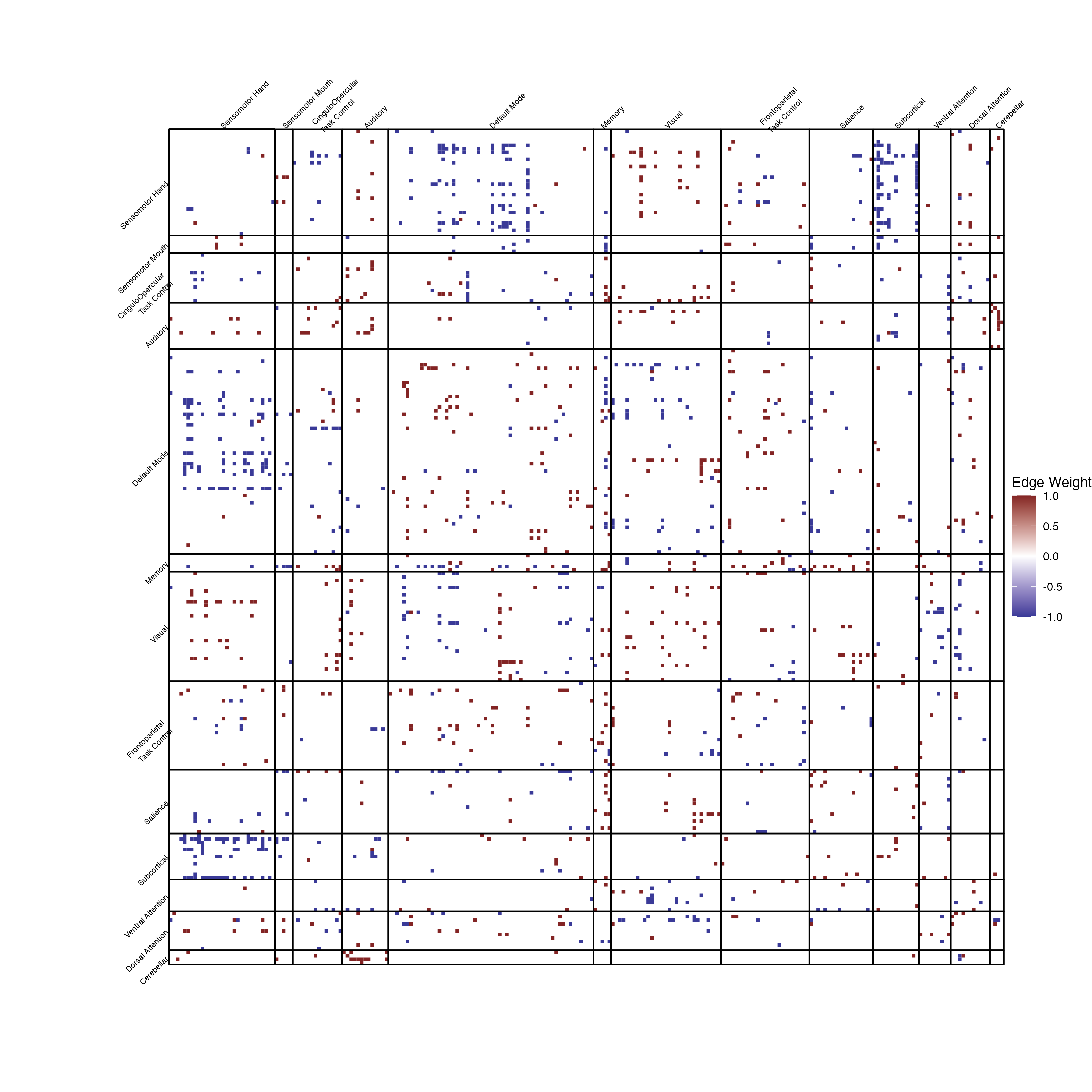}
  \caption{
    Edges selected by CPM, colored by the sign of their association with GCA.
    Solid lines depict boundaries of the Power parcellation.
  }
  \label{fig:cpm-g}
\end{figure}

\begin{figure}[ht]
  \centering
  \includegraphics[width=1.0\textwidth]{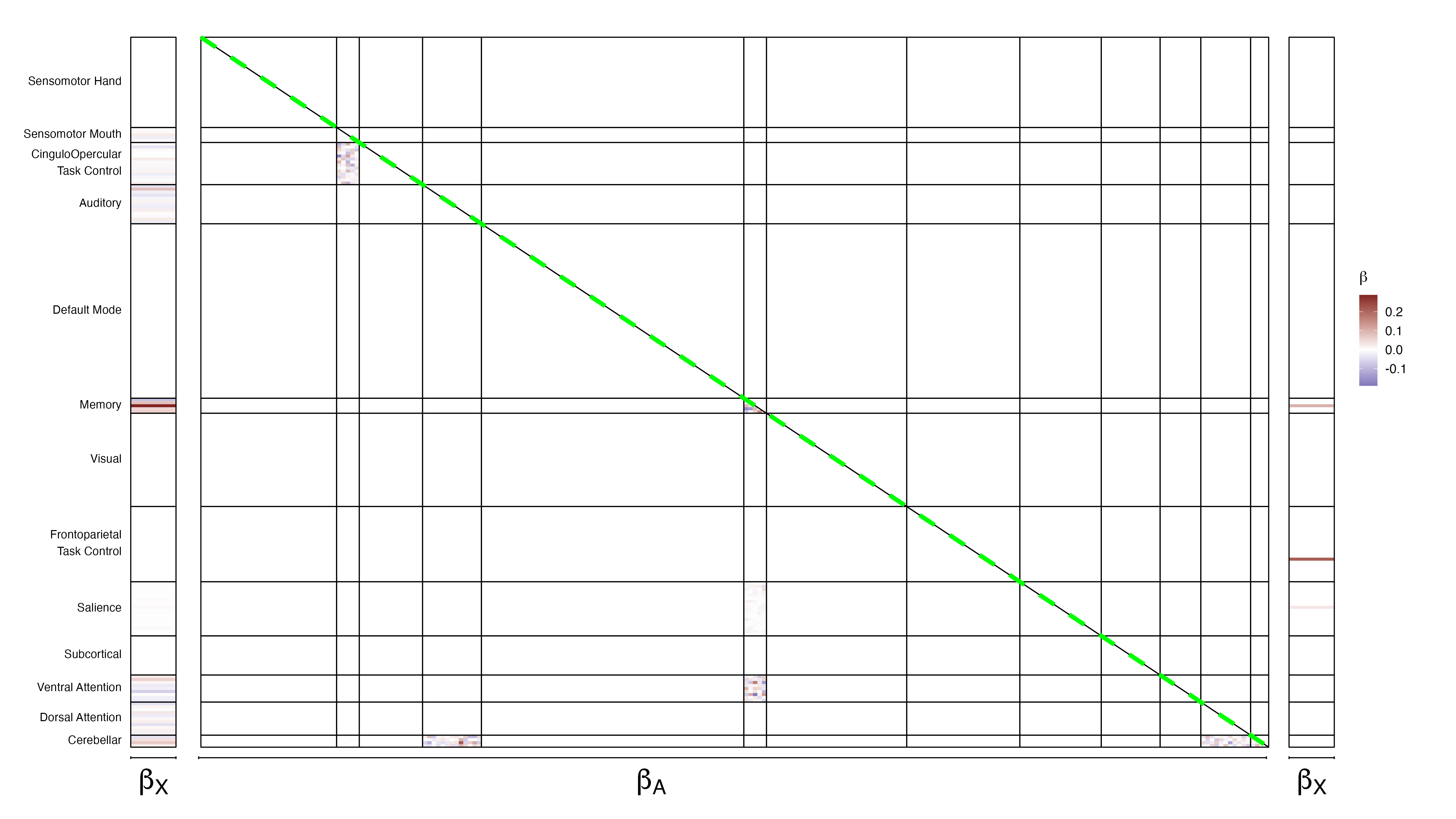}
  \caption{
    Visualization of $\V{\beta}$ coefficients with PMAT as response.
    Coefficients from NetCov with EBG are presented at left ($\M{\beta}_{\M{X}}$) and on the lower triangle ($\M{\beta}_{\M{A}}$).
    Coefficients from LASSO are presented at right ($\M{\beta}_{\M{X}}$) and on the upper triangle ($\M{\beta}_{\M{A}}$).
    Solid lines depict boundaries of the Power parcellation.
  }
  \label{fig:hcp-coef-pmat}
\end{figure}

\begin{figure}[ht]
  \centering
  \includegraphics[width=1.0\textwidth]{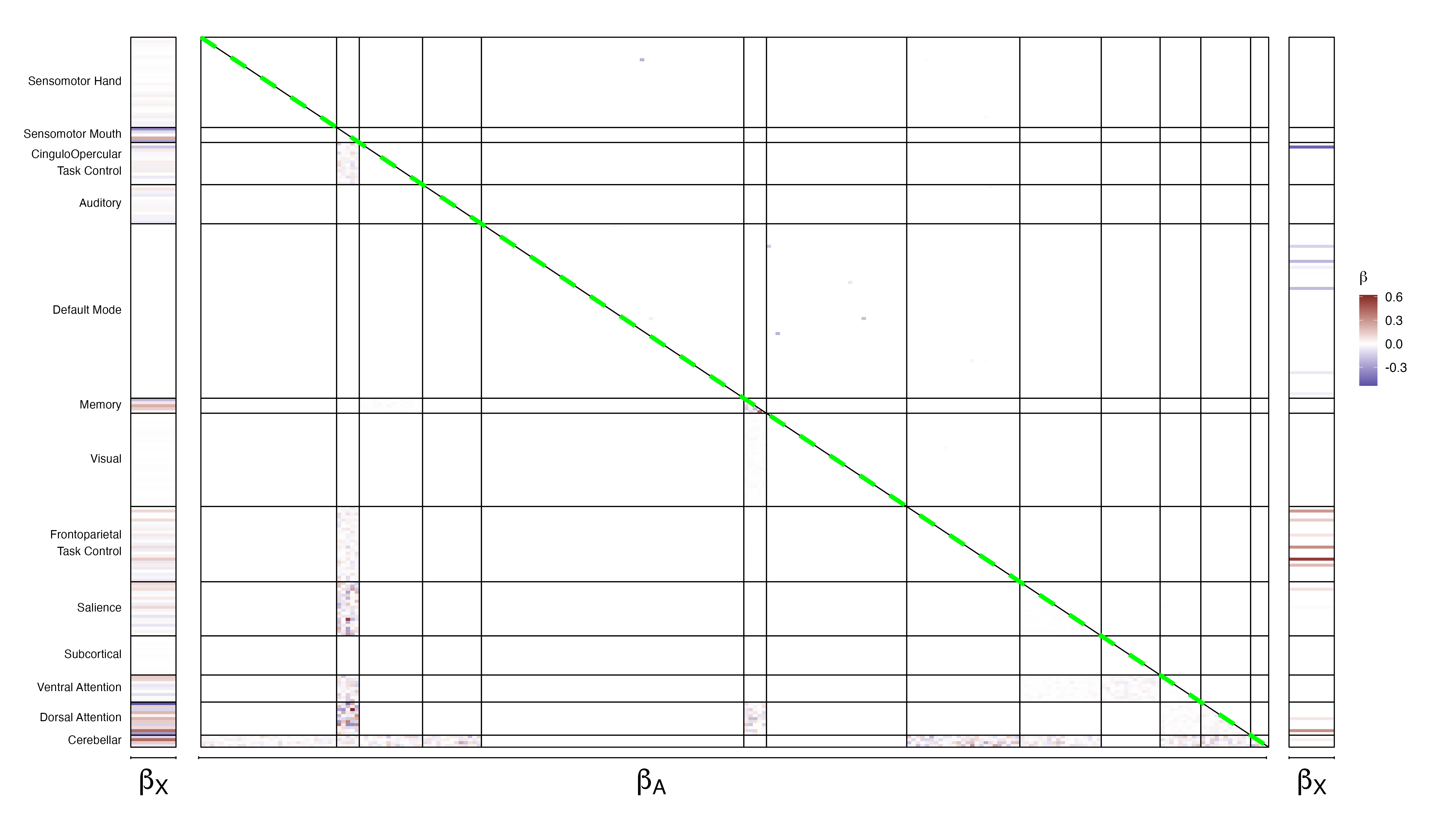}
  \caption{
    Visualization of $\V{\beta}$ coefficients with Working Memory as response.
    Coefficients from NetCov with EBG are presented at left ($\M{\beta}_{\M{X}}$) and on the lower triangle ($\M{\beta}_{\M{A}}$).
    Coefficients from LASSO are presented at right ($\M{\beta}_{\M{X}}$) and on the upper triangle ($\M{\beta}_{\M{A}}$).
    Solid lines depict boundaries of the Power parcellation.
  }
  \label{fig:hcp-coef-wm}
\end{figure}

\begin{figure}[ht]
  \centering
  \includegraphics[width=1.0\textwidth]{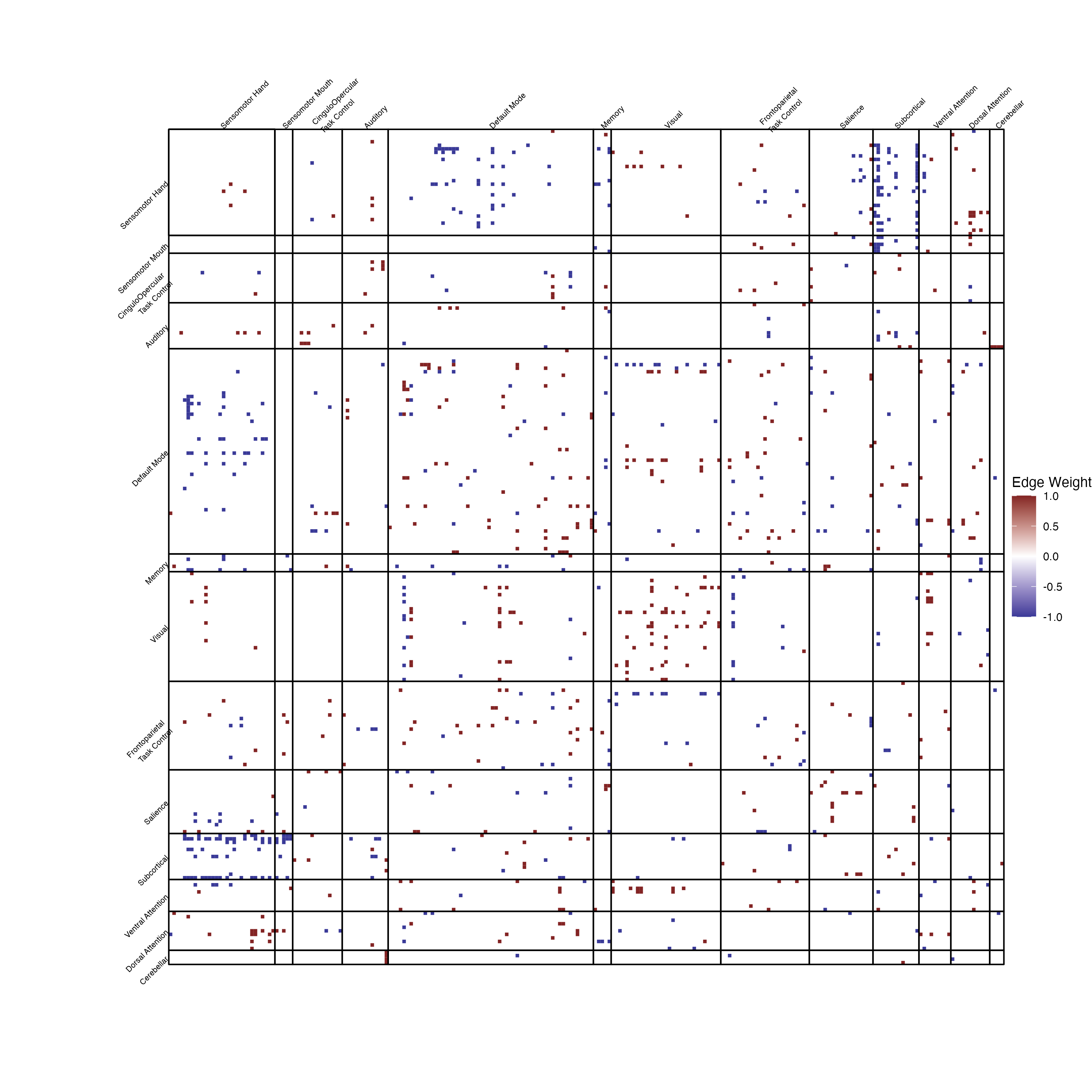}
  \caption{
    Edges selected by CPM, colored by the sign of their association with PMAT.
    Solid lines depict boundaries of the Power parcellation.
}
  \label{fig:cpm-pmat}
\end{figure}

\begin{figure}[ht]
  \centering
  \includegraphics[width=1.0\textwidth]{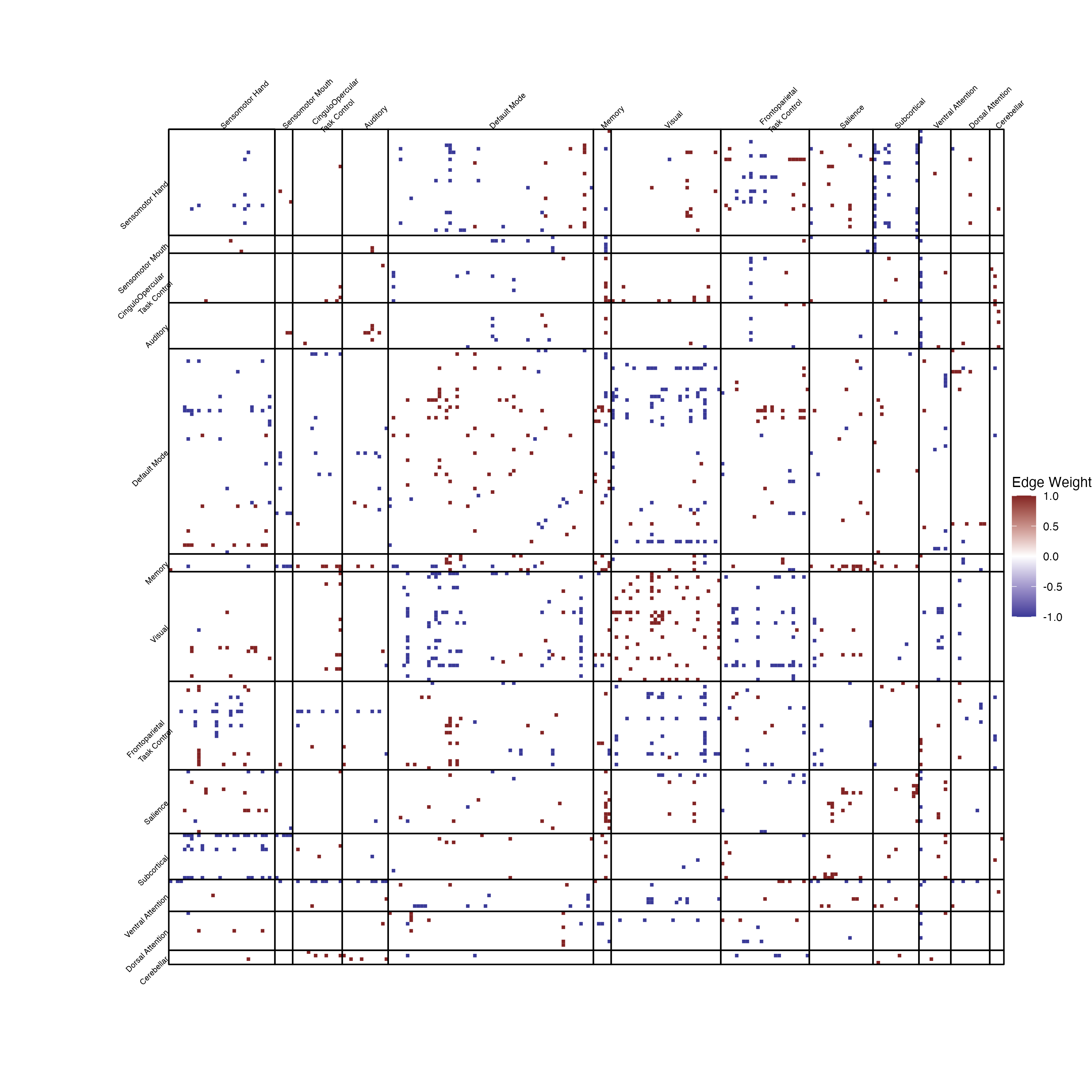}
  \caption{
    Edges selected by CPM, colored by the sign of their association with Working Memory.
    Solid lines depict boundaries of the Power parcellation.
}
  \label{fig:cpm-wm}
\end{figure}

In \cref{fig:cpm-g}, we present the edges selected by CPM for predicting GCA.
We present the estimated coefficients for $\M{\beta}_{\M{A}}$ and $\M{\beta}_{\M{X}}$ from both NetCov (EBG grouping) and LASSO associated with PMAT and Working Memory in \cref{fig:hcp-coef-pmat,fig:hcp-coef-wm}.
In \cref{fig:cpm-pmat,fig:cpm-wm}, we depict edges selected by CPM for PMAT and Working Memory.

\end{document}